\documentclass{aa}  

\usepackage{graphicx}
\usepackage{booktabs}
\usepackage{txfonts}

\usepackage{subfig}
\usepackage{afterpage}
\usepackage{float}
\usepackage{graphicx}
\usepackage{adjustbox}
\usepackage{gensymb}
\usepackage{enumitem}
\usepackage{pifont}
\usepackage[nolist]{acronym}

\begin{acronym}
        \acro{cme}[CME]{coronal mass ejection}
        \acroplural{cme}[CMEs]{coronal mass ejections}
        \acro{icme}[ICME]{interplanetary coronal mass ejection}
        \acro{gle}[GLE]{ground level enhancement}
        \acroplural{gle}[GLEs]{ground level enhancements}
        \acroplural{icme}[ICMEs]{interplanetary coronal mass ejections}
        \acro{sep}[SEP]{solar energetic particle}
        \acroplural{sep}[SEPs]{solar energetic particles}
        \acro{sir}[SIR]{stream interaction region}
        \acroplural{sir}[SIRs]{stream interaction regions}
        \acro{cir}[CIR]{corotating interaction region}
        \acroplural{cir}[CIRs]{corotating interaction regions}  
        \acro{gcs}[GCS]{graduated cylindrical shell}
        \acro{sta}[STA]{STEREO~A}
        \acro{stb}[STB]{STEREO~B}
        \acro{ar}[AR]{active region}
        \acroplural{ar}[ARs]{active regions}
\end{acronym}

\begin{document}

        \title{Long-lasting injection of solar energetic electrons into the heliosphere}
        
   \author{N. Dresing\inst{1}
          \and
          R. G\'omez-Herrero\inst{2}
          \and
          B. Heber\inst{1}
          \and
          A. Klassen\inst{1}
          \and
          M. Temmer\inst{3} 
          \and
          A. Veronig\inst{3}
          }

   \institute{Institut f\"ur Experimentelle und Angewandte Physik, University of Kiel, Germany\\
   \email{dresing@physik.uni-kiel.de}   
         \and
             Space Research Group, Dpto. de F\'isica y Matem\'aticas, University of Alcal\'a, Spain\\   
          \and
             Institute of Physics/Kanzelhöhe Observatory, University of Graz, Austria\\
             }

   \date{}

 
  \abstract
   {The main sources of \ac{sep} events are solar flares and shocks driven by \acp{cme}. While it is generally accepted that energetic protons can be accelerated by shocks, whether or not these shocks can also efficiently accelerate solar energetic electrons is still debated.\\
        In this study we present observations of the extremely widespread SEP event of 26 Dec 2013.
    To the knowledge of the authors, this is the widest longitudinal SEP distribution ever observed together with unusually long-lasting energetic electron anisotropies at all observer positions. Further striking features of the event are long-lasting SEP intensity increases, two distinct SEP components with the second component mainly consisting of high-energy particles, a complex associated coronal activity including a pronounced signature of a shock in radio type-II observations, and the interaction of two \acp{cme} early in the event.
        }
   {The observations require a prolonged injection scenario not only for protons but also for electrons. We therefore analyze the data comprehensively to characterize the possible role of the shock for the electron event.}
   {Remote-sensing observations of the complex solar activity are combined with in-situ measurements of the particle event. We also apply a Graduated Cylindrical Shell (GCS) model to the coronagraph observations of the two associated \acp{cme} to analyze their interaction.}
   {We find that the shock alone is likely not responsible for this extremely wide SEP event. Therefore we propose a scenario of trapped  energetic particles inside the \ac{cme}-\ac{cme} interaction region which undergo further acceleration due to the shock propagating through this region, stochastic acceleration, or ongoing reconnection processes inside the interaction region. The origin of the second component of the \ac{sep} event is likely caused by a sudden opening of the particle trap.}
   {}
        

   \maketitle
%

\acresetall

\section{Introduction}\label{sec:intro}
\ac{sep} events mainly consist of electrons and protons with small amounts of heavier ions.
Two phenomena are considered to be the main accelerators of these particles: Solar flares and shocks driven by \acp{cme}.
The widely-used classification by \citet{Reames1999} distinguishes between impulsive, that is, flare accelerated events, and gradual, that is, shock-associated events.
Impulsive events show enrichments of $^3$He, heavy elements such as Fe, and electrons, while gradual events are proton- and ion-rich showing compositions more similar to coronal and solar wind material.
Naturally, the extended shock front builds a larger acceleration region producing a larger SEP spread in the inner heliosphere.
However, a significant number of electron events with longitudinal spreads much larger than the expected widths of impulsive events \citep[$\sigma\sim20$ degrees,][]{Reames1999} have been observed \citep{Wibberenz2006}. 
Thanks to the two STEREO spacecraft it was possible to detect so-called widespread events \citep{Dresing2012, Lario2013, Dresing2014, Gomez-Herrero2015} where SEPs including electrons were distributed all around the Sun.
Also, unexpectedly wide $^3$He events were observed with the STEREO spacecraft \citep{Wiedenbeck2013}.
The important processes for these extraordinarily wide particle spreads are a matter of debate.\\
Some authors have suggested that strong perpendicular diffusion might explain the wide particle spreads \citep{Dresing2012, Droege2014, Droege2016, Strauss2017}.
Others suspect a large shock to be the driver \citep{Richardson2014, Lario2014}.
Especially in the case of electrons it is not clear if a shock is able to efficiently accelerate electrons during SEP events providing an extended source region.
No in-situ measurements close to the acceleration sites near the Sun exist and the likely mixing of flare accelerated and possibly shock accelerated electrons makes it very difficult to distinguish the populations when measured at spacecraft.
While interplanetary shocks, which can be measured in-situ at a spacecraft, have shown to be very inefficient in accelerating electrons ($1\%$ for $<100$\,keV electrons \citep{Tsurutani1985, Dresing2016b}, this might be different close to the Sun where the Alfv\'en speed is larger  \citep{Mann1999, Gopalswamy2001b} and a greater turbulence and seed-particle population might be present.
Especially preceding \acp{cme} can have a preconditioning effect on the ambient medium and have been found to be connected with higher fluxes of \ac{sep} events \citep{Lugaz2017}.\\
The existence of galactic cosmic ray electrons suggests that shocks are capable of accelerating electrons to high energies \citep{Aharonian2005}. 
Furthermore, the observation of "herringbone" structures in type-II radio bursts during solar events indicates the presence of shock-accelerated electrons of at least a few tens of keV \citep{Mann2005}.
However, whether or not coronal or \ac{cme}-driven shocks can accelerate electrons up to hundreds of keV or a few MeV, such as what is observed during SEP events, is still uncertain.      
On the one hand, \cite{Haggerty2002} explain the onset delays between the inferred injection time of near-relativistic electrons and the associated flare/type III times with the shock being the electron source.
On the other hand, \citet{Maia2004} suggest that an ongoing reconnection process in the corona driven by the uplifting \ac{cme} accelerates the electrons and may account for the delayed onsets. 
Although some authors argue     that coronal shocks \citep{Klassen2002} or standing fast mode shocks involved in flares \citep{Mann2009}, are capable of accelerating electrons up to relativistic energies, an unambiguous distinction between shock acceleration or particle release due to the shock has not yet been made.
These delays could, however, also be caused by turbulence and its resulting transport effects, such as scattering or an effective increase of the connecting field line length.\\
Detailed case studies of extreme events, like widespread events, may help to disentangle the main physical processes involved.
The event presented in this study is a widespread electron (and proton) event which was observed on 26 Dec 2013 by the two STEREO spacecraft, separated by 59 degrees at that time, and by the L1-spacecraft SOHO, ACE, and Wind, separated by 150 and 151 degrees with respect to \ac{sta} \ac{stb}, respectively.
The event is associated with complex solar activity and a prominent type-II radio burst extending far into the interplanetary (IP) medium.
The electron intensities show long-lasting rising phases and unusually long-lasting anisotropies observed at all three positions.
The onset delays between the spacecraft are small, even at L1, being poorly connected to the solar event.
However, all energetic particle onsets at the three observers show a delay of at least $\sim$30 minutes with respect to the assumed flare injection time.
A long-lasting and spatially extended injection is therefore required to explain the observations.
Although the \ac{cme}-driven shock seems to be a good candidate, we carefully analyze the event and find that the shock alone does not explain all the characteristics of the event; a complex scenario also involving particle trapping is more likely.\\
In Sect. \ref{sec:obs} we first discuss the remote-sensing observations of the associated complex solar event including the interaction of two \acp{cme} (Sect. \ref{sec:cme_interaction}). 
Then we present the energetic particle observations at all three observers (Sect. \ref{sec:sep_obs}) followed by the interplanetary context (Sect. \ref{sec:IPcontext}).
Afterwards we determine the longitudinal spread of the analyzed event in Sect. \ref{sec:spread} and finally we discuss the possible source of the SEP event in Sect. \ref{sec:second_comp}.
\section{Instrumentation}\label{sec:instr}
The SEP observations at the STEREO spacecraft are provided by the instruments of the IMPACT suite \citep{Luhmann2007}: the High Energy Telescope \cite[HET,][]{Rosenvinge2008}, the Low Energy Telescope \cite[LET,][]{Mewaldt2007}, and the Solar Electron and Proton Telescope \cite[SEPT,][]{Muller-Mellin2007}. 
Directional SEP measurements can be obtained only from LET (protons from 1.8-15\,MeV) and SEPT (electrons from 30-400\,keV, protons from 60-7000\,keV).
Solar wind plasma and magnetic field observations are provided by the PLASTIC \citep{Galvin2008} and MAG \citep{Acuna2007} instruments, respectively.
The SECCHI instrument suite \citep{Howard2008} contains the remote-sensing instrumentation of the STEREO spacecraft. 
Observations of the coronagraphs COR1 and COR2, as well as images of the EUV cameras \citep[EUVI,][]{Wuelser2004} were used in this study.
Radio observations are provided by the SWAVES instruments \citep{Bougeret2008}.
The STEREO observations of this study are complemented by observations from the Earth's point of view:
We use energetic particle measurements taken by the EPHIN instrument \citep{Muller-Mellin1995} aboard SOHO, by EPAM \citep{Gold1998} aboard ACE and by the 3DP detector \citep{Lin1995} aboard Wind.
Solar wind plasma data at the Lagrangian point L1 were provided by ACE/SWEPAM \citep{McComas1998}.
The WAVES instrument aboard WIND \citep{Bougeret1995} provides radio measurements at this position.
EUV imaging from SDO's AIA instrument \citep{Lemen2012} and the coronagraph LASCO \citep{Brueckner1995} aboard SOHO complete the 360 degree remote-sensing set.
%
\section{Observations}\label{sec:obs}
\subsection{Remote-sensing observations}\label{sec:remote}
\begin{figure*}[h!]
        \centering
        \begin{minipage}{.33\linewidth}
                \centering
                \subfloat[]{\includegraphics[width=1.01\textwidth, clip=true, trim = 0mm 0mm 0mm -4mm]{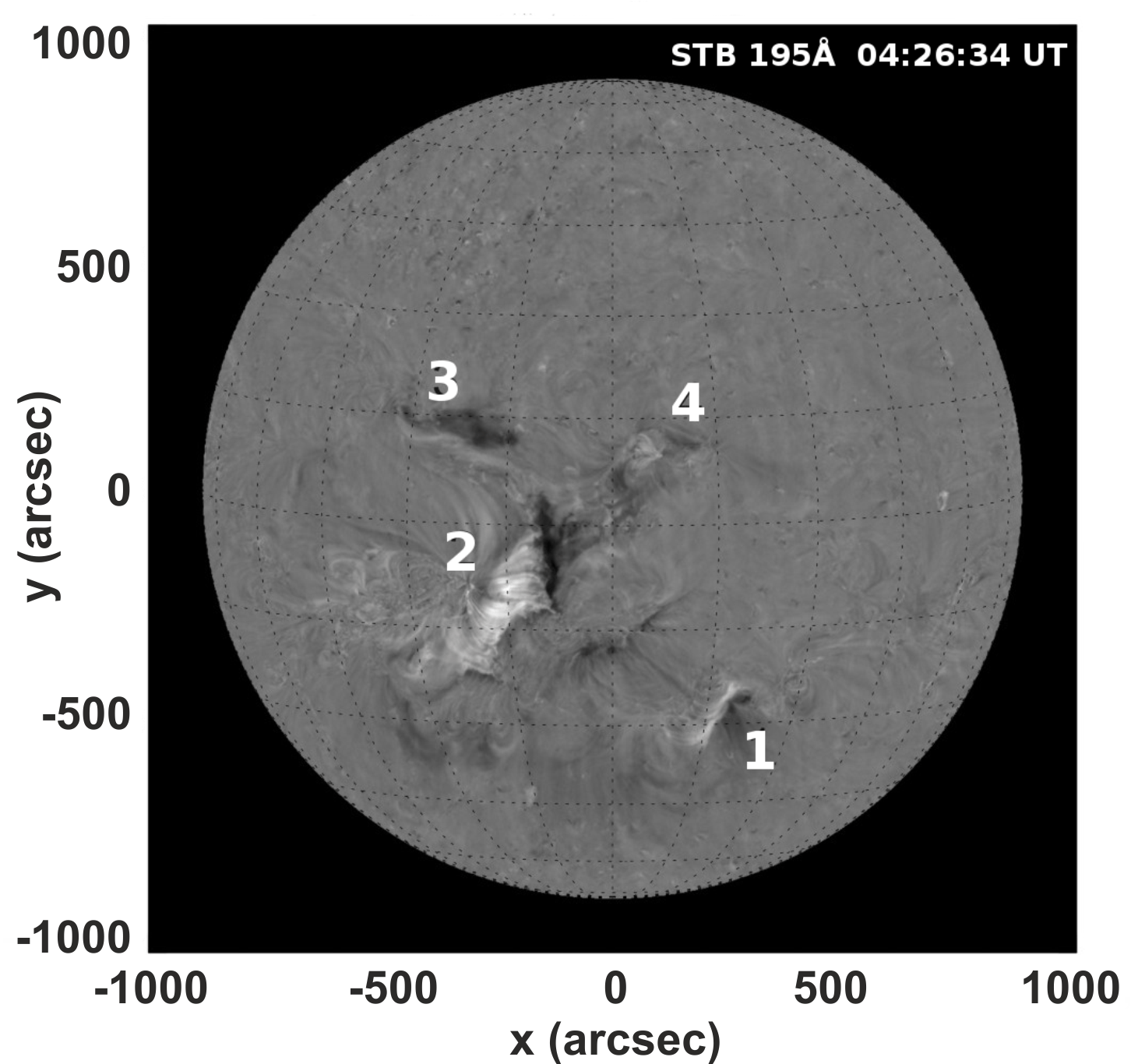}}
        \end{minipage}
        \begin{minipage}{.33\linewidth}
                \centering
                \subfloat[]{\includegraphics[width=0.89\textwidth, clip=true, trim = 0mm 0mm 0mm 0mm]{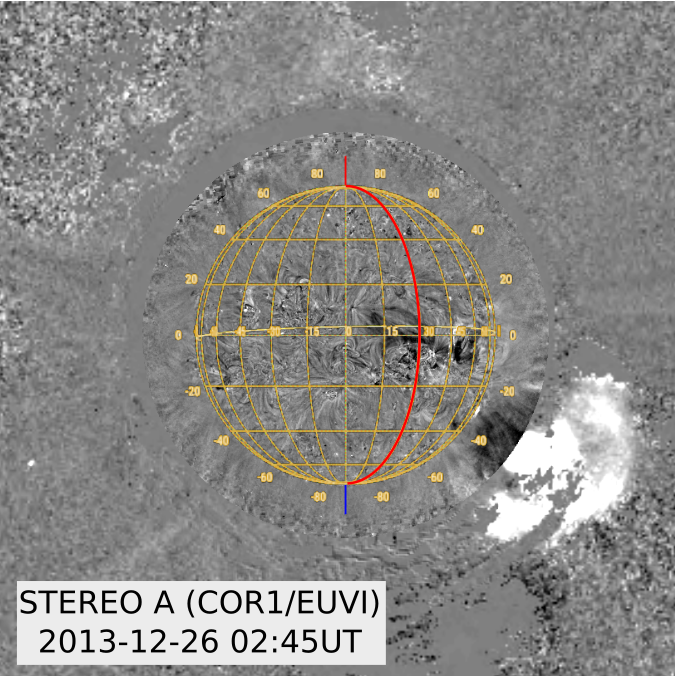}}
        \end{minipage}%
        \begin{minipage}{.33\linewidth}
                \centering
                \subfloat[]{\includegraphics[width=0.939\textwidth, clip=true, trim = 0mm -5.5mm 0mm -6mm]{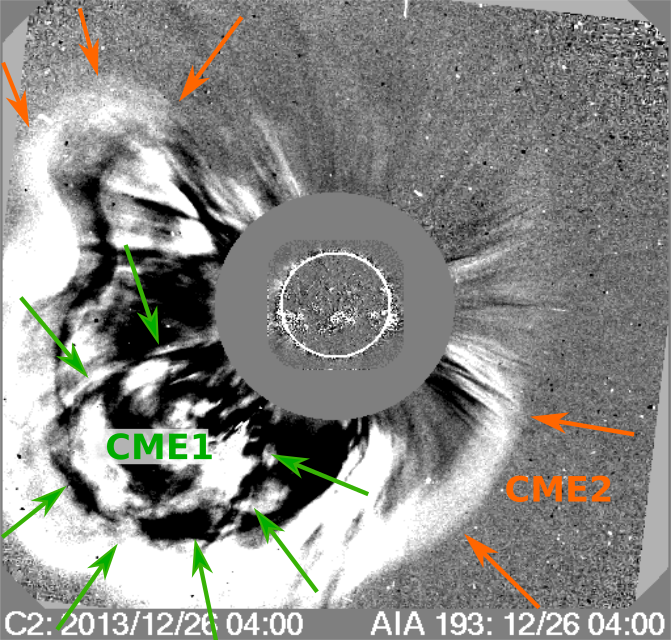}}
        \end{minipage}
        \caption{(a) \ac{stb}/EUVI 195{\AA} base difference image at 4:27\,UT. The grid lines correspond to a 15 degree Carrington grid. The four regions of activity are marked by numbers. (b) Combined EUVI and COR1 difference image taken by \ac{sta} at 2:35\,UT. (c) Combined EUV and coronagraph difference images observed by SOHO/C2 and SDO/AIA at 4:00\,UT. \ac{cme}1 is marked by green arrows for the time when it is overtaken by \ac{cme}2 (orange arrows).} \label{fig:constellation_EUV}
\end{figure*}%
\begin{figure}
        \includegraphics[width=0.45\textwidth, clip=true, trim = 0mm 0mm 0mm 0mm]{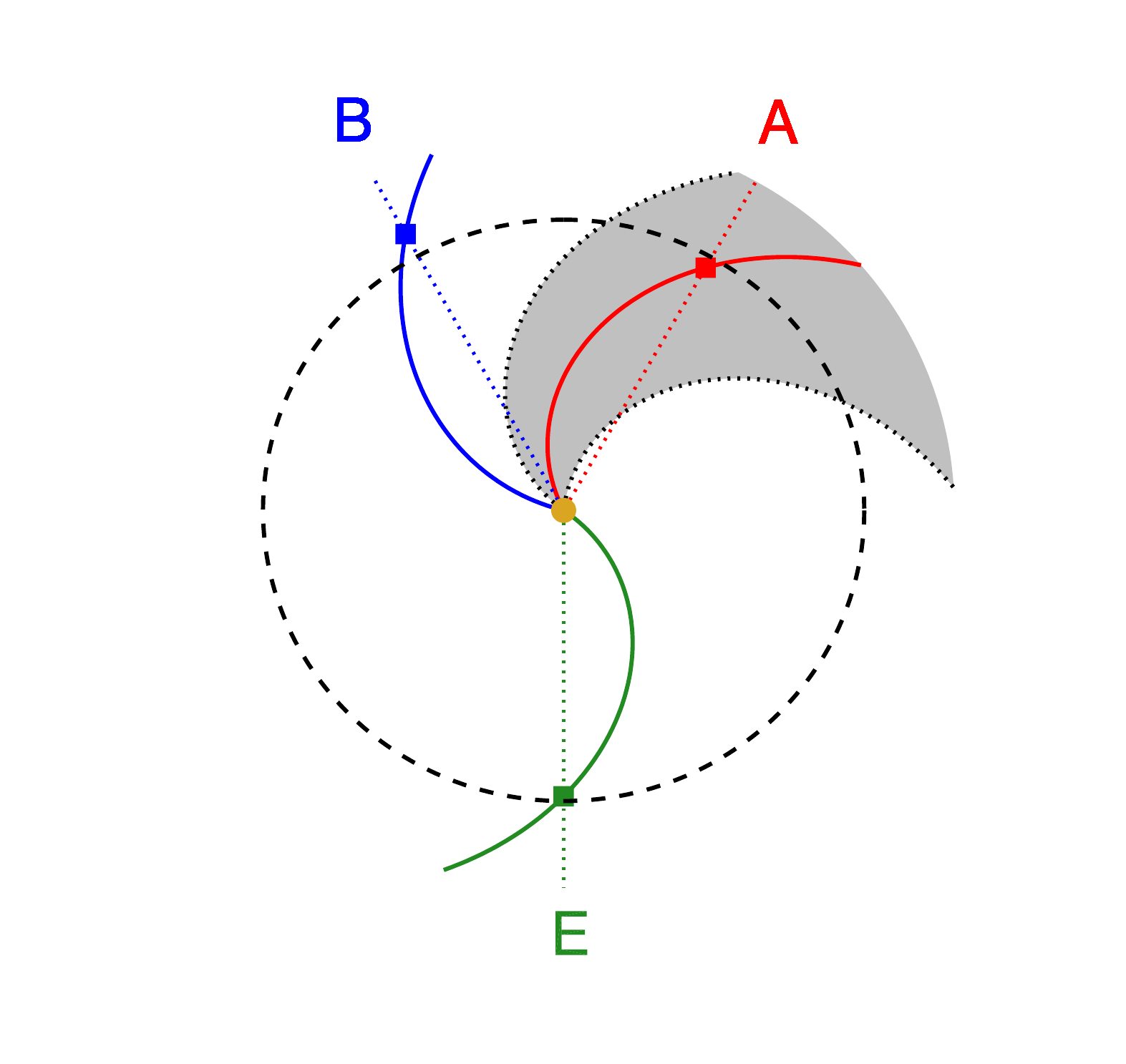}
        \caption{Longitudinal constellation of the Earth (E) and the two STEREO spacecraft (A,B) in the ecliptic plane with respect to the region of activity at the Sun (gray shaded sector). The colored spirals represent the magnetic field lines connecting the observers with the Sun taking into account the measured solar wind speed.}
        \label{fig:constellation}
\end{figure}
%
%
\begin{figure}[h!]
        \centering
        \includegraphics[width=0.4\textwidth, clip=true, trim = 0mm 0mm 0mm 0mm]{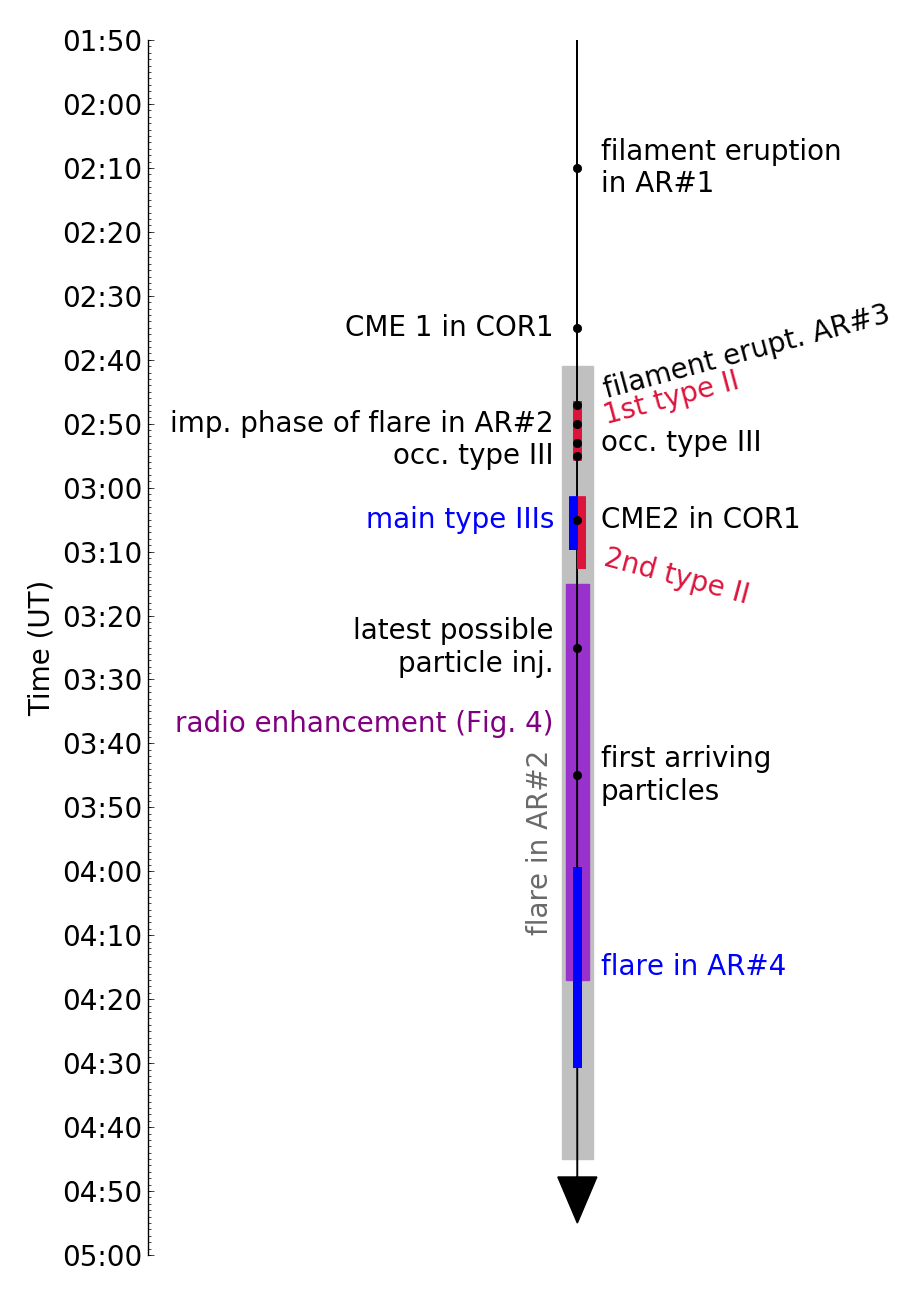}
        \caption{Timeline of the early solar event according to Table \ref{table:timeline}.} \label{fig:timeline}
\end{figure}%
\begin{figure}[h!]
        \centering
        \includegraphics[width=0.5\textwidth, clip=true, trim = 0mm 0mm 0mm 0mm]{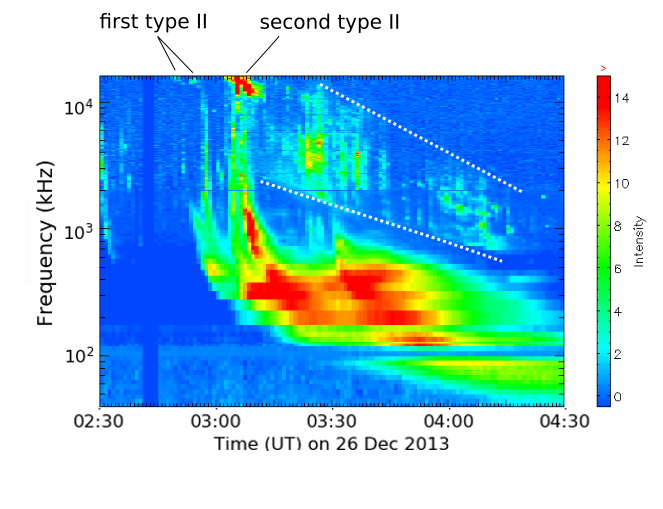}
        \caption{Dynamic radio spectrum observed by \ac{stb}/SWAVES during 2:30-4:40\,UT on 26 Dec 2013.} \label{fig:radio2}
\end{figure}%
\begin{figure}[h!]
\centering
\includegraphics[width=0.45\textwidth, clip=true, trim = 0mm 26mm 0mm 0mm]{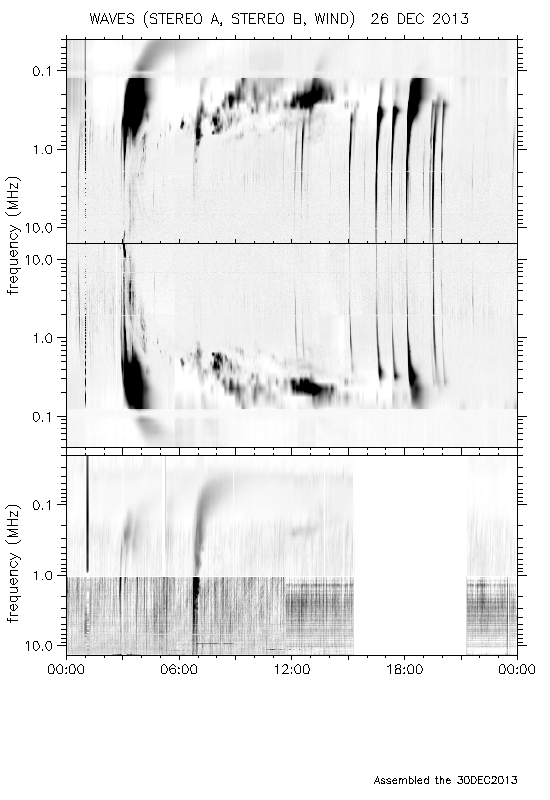}
  \caption{Dynamic radio spectra observed at \ac{sta} (top), \ac{stb} (center, frequency-axis reversed), and Wind (bottom) by courtesy of \protect\url{http://secchirh.obspm.fr/select.php}.} \label{fig:radio}
\end{figure}%
%
%
%
\begin{table*}[h!]
        \centering
        \caption{Time-line of the solar phenomena and energetic electron onset times associated with the 26 Dec 2013 SEP event. Observers of radio features in brackets denote a high-frequency occultation for that viewpoint.}
        \begin{tabular}{lll}
                \toprule
                Time (UT)      & Properties                                                    & Observer \\
                \midrule
                2:10           & filament eruption (region \#1) associated to \ac{cme}1             & \ac{sta}, \ac{stb} /EUVI\\
                2:30           & \ac{cme}1 at 1\,R$_S$                                              & \ac{sta}, \ac{stb} \\
                2:35           & \ac{cme}1 first appearance                                         & \ac{sta}/COR1 \\
        2:41 -  4:45   & main flare (region \#2)                                            & \ac{sta}, \ac{stb}\\
                2:47 - 2:55    & 1st (faint) type II burst                                     & \ac{stb}/SWAVES \\
                2:47           & filament eruption (region \#3) and interaction with northern CH & \ac{stb}/EUVI \\
                2:50           & impulsive phase of main flare                                 & \ac{sta}, \ac{stb} /EUVI\\
        2:53           & partly occulted type III burst                                & \ac{sta}, \ac{stb}, Wind\\
        2:55           & partly occulted type III burst                                & \ac{sta}, \ac{stb}, Wind\\
                3:02 - 3:12    & 2nd (main) type II burst                                      & \ac{stb}, (\ac{sta}) \\
                3:02 - 3:09    & main type III bursts                                          & \ac{sta}, \ac{stb}, (Wind/WAVES)\\
                3:05           & \ac{cme}2 first appearance at 2.7\,R$_{\mbox{s}}$                                         & \ac{sta}/COR1 \\
            3:15 - 4:17    & drifting type II-like structure (see Fig. \ref{fig:radio2})   & \ac{stb}/SWAVES \\
            4:00 - 4:30    & flare in region \#4                                           & \ac{stb}/EUVI\\     
                &&\\
                3:25             & latest possible 55-105\,keV electron injection (according to    & \\
                                 & onset time assuming a nominal travel time along the Parker spiral)               & \\
                3:45\,$\pm$1min  & 55-105\,keV electron onset                                    & \ac{sta}/SEPT \\
                3:53\,$\pm$5min  & 55-105\,keV electron onset                                    & \ac{stb}/SEPT \\
                4:10\,$\pm$15min & 62-103\,keV electron onset                                    & ACE/EPAM \\
                &&\\
                4:07\,$\pm$15min & 0.7-2.8\,MeV electron onset                                   & \ac{sta}/HET \\
                4:22\,$\pm$15min & 0.7-2.8\,MeV electron onset                                   & \ac{stb}/HET \\
                4:30\,$\pm$20min & 0.7-3.0\,MeV electron onset                                   & SOHO/EPHIN \\
                &&\\
                6:45           & type III burst                                                & Wind/WAVES \\
                6:53           & C2.2 flare at W28                                             & GOES, SDO \\
                7:05           & \ac{cme}3 first appearance                                         & \ac{sta}/COR1 \\
                7:15 - 8:30        & SEP onsets of the second component                                                 & \ac{sta}, \ac{stb} /HET, SOHO/EPHIN,ERNE \\
                \bottomrule
        \end{tabular}
        \label{table:timeline}
\end{table*}
The solar event on 26 Dec 2013 is a very complex event involving four \acp{ar} of distinct activity at the Sun (numbered from one to four in Fig. \ref{fig:constellation_EUV} (a)). 
In total it has a spatial extent of more than 60 degrees in latitude and longitude centered at the central meridian as seen from \ac{stb}.
This extent makes it difficult to provide a single coordinate which can be assumed to be the injection site of the flare-accelerated SEPs.
Figure \ref{fig:constellation} shows the longitudinal constellation of the two STEREO spacecraft (\ac{stb} in blue and \ac{sta} in red) and the Earth (green) on 26 Dec 2013.
The gray sector indicates the 60-degree-wide longitudinal range magnetically connected to the suspected solar source region longitudes according to the region of activity involved at the Sun. 
While \ac{sta} is magnetically connected to that sector, the magnetic footpoint of \ac{stb} lies outside towards western longitudes.
The Earth is situated and connected to the backside of this sector of activity at the Sun.\\
Figure \ref{fig:constellation_EUV} (a) shows an extreme ultraviolet (EUV) base difference image at 195\,{\AA} observed by \ac{stb} at 4:27\,UT.
The different numbers indicate the \acp{ar} ordered by the sequence of activity. 
A time-line of the coronal events can be found in Table \ref{table:timeline} which is also illustrated in Fig. \ref{fig:timeline}.
At 2:10\,UT \ac{stb} (and \ac{sta}) observe a filament eruption in region \#1 towards south/west.
This filament eruption is associated to a \ac{cme} listed in the LASCO catalog\footnote{\url{https://cdaw.gsfc.nasa.gov/CME_list/UNIVERSAL/2013_12/univ2013_12.html}} with a projected speed of 1022\,km/s, a position angle (PA) of 122 degrees, and an angular width of $>171$ degrees.
The kinematics of the \ac{cme} suggest that the filament eruption would reach 1\,R$_{\rm S}$ (above the solar surface) at 2:30\,UT which agrees with the observations at \ac{stb} where the \ac{cme} appears in COR1 (1.4\,R$_{\rm S}$) at that time.
We note that all heights provided in R$_{\rm S}$ in this paper denote heights above the solar surface.
Figure \ref{fig:constellation_EUV} (b) shows a combined EUVI and COR1 difference image observed by \ac{sta} showing the early \ac{cme} in the south/east, henceforth referred to as \ac{cme}1.\\
The main event is a large two-ribbon flare in region \#2, located at E17S12 as seen from \ac{stb} point of view (cf. Fig. \ref{fig:constellation_EUV} (a)) and starting at 2:41\,UT, which likely triggers the later activity in regions \#3 and \#4.
According to the statistical relation found by \citet{Nitta2013}, we estimate the GOES class of this flare as M7 utilizing the disk-integrated emission change in 195\,{\AA} observed by \ac{stb} (not shown here).
The flare is accompanied by an EIT wave propagating mainly towards west and south (not shown here). 
The propagation towards north is blocked by a coronal hole (CH) and towards east by an AR.\\
This event is also accompanied by another large \ac{cme} listed in the LASCO catalog as a halo \ac{cme} with a speed of 1336\,km/s.
Based on the propagation direction of the EIT wave the \ac{cme} is likely deflected towards south/west making an interaction with \ac{cme}1 likely (see Sect. \ref{sec:cme_interaction}).
Figure \ref{fig:constellation_EUV} (c) shows the observations of this \ac{cme} (henceforth \ac{cme}2) from the Earth's point of view.
\ac{cme}1 (marked by arrows in Fig. \ref{fig:constellation_EUV} (c)) clearly posed an obstacle for \ac{cme}2 which is reflected in the disturbed front of \ac{cme}2, suggesting interaction between both \acp{cme}.
A discussion on that interaction follows in Sect. \ref{sec:cme_interaction}.\\
The activity in region \#2 triggers region \#3, close to the northern polar CH, where a plasma loop moving towards the northern CH leads to an enlargement of the CH at its southern boundary.
A filament eruption is then observed at 2:47\,UT in region \#3.
The loop connecting regions \#2 and \#3 is transequatorial.
Finally, a small flare in region \#4 (cf. Fig. \ref{fig:constellation_EUV} (a)) is observed from 4:00 to 4:30\,UT.
The 195\,{\AA} observations at \ac{stb} reveal that there is a connection between the main event in region \#2 and the flare in region \#4.\\
The solar event was associated with type-II and type-III radio bursts best observed by \ac{stb}/SWAVES.
Two short-lived type-II bursts, signatures of shock waves in the corona, appeared at frequencies above 10\,MHz.
We note that no ground-based radio signals at higher frequencies were recorded due to the backside location of the event as seen from Earth. 
The first (faint) type-II burst occurred between 02:47 and 2:55\,UT followed by partly occulted type-III radio bursts
(see Fig. \ref{fig:radio2} showing the radio observations by \ac{stb} during the early phase of the event).
The second main type-II burst appeared between 3:02 and 3:12\,UT, a few minutes after the first one ceased,
and simultaneously with the main group of type-III bursts, which were observed by \ac{sta} and \ac{stb} but not at Wind.\\
Figure \ref{fig:radio} shows the dynamic radio spectra of the whole day of 26 Dec as detected on board \ac{sta} (top), \ac{stb} (middle, y-axis reversed), and Wind (bottom).
While \ac{stb} observed the high-frequency part coming from deeper regions in the corona, this part is slightly occulted for \ac{sta} and all of the radio type-III bursts are only barely visible at Wind.
%
\begin{figure*}[h!]
        \centering
        \includegraphics[width=0.91\textwidth, clip=true, trim = 0mm 0mm 0mm 0mm]{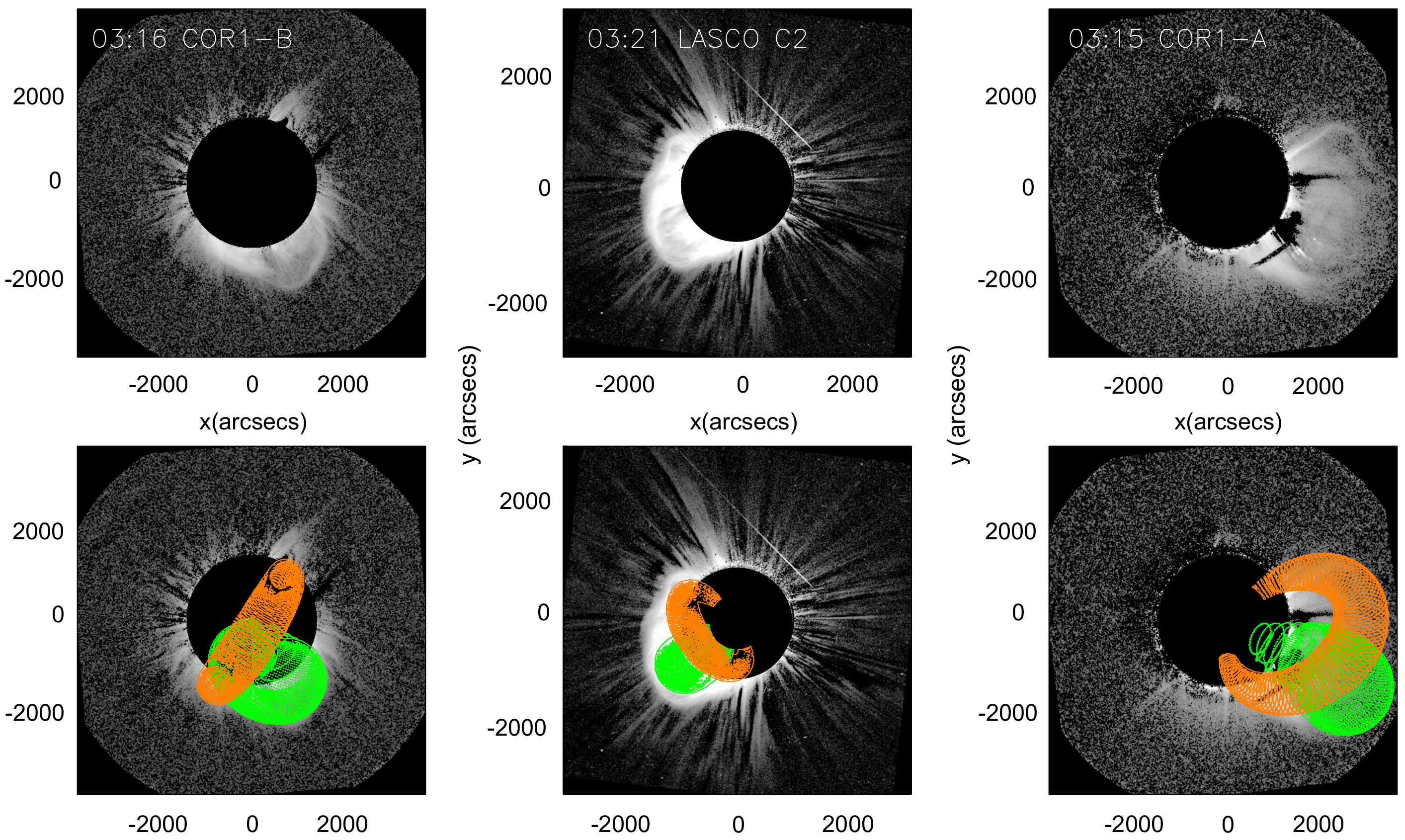}
        \caption{Top: Difference images of coronagraph observations at \ac{stb} (left), SOHO (center), and \ac{sta} (right). Bottom: The same difference images overlayed with the results of a GCS model reproducing the flux ropes of \ac{cme}1 (green) and \ac{cme}2 (orange).} \label{fig:gcs}
\end{figure*}
\begin{figure}[h!]
        \centering
        \includegraphics[width=0.48\textwidth, clip=true, trim = 0mm 0mm 0mm 0mm]{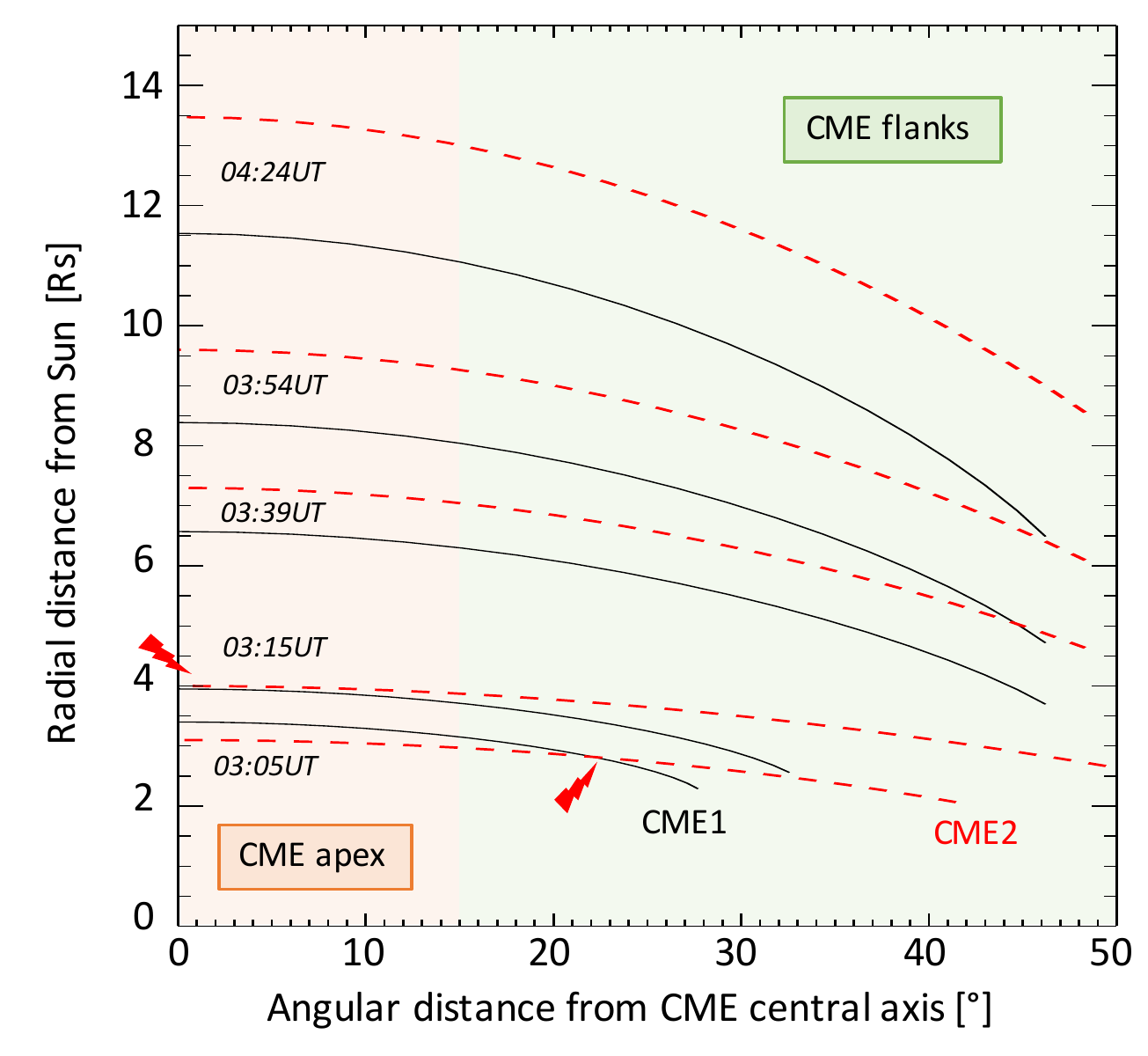}
        \caption{GCS model results (cf. Fig. \ref{fig:gcs}) showing the heights of the reconstructed fronts of \ac{cme}1 (solid black) and \ac{cme}2 (dashed red) as a function of their angular widths at five time steps according to the COR1 cadence. We defined the apex to cover an angle of 15 degrees from the central axis, and the flanks to cover the angles larger than 15 degrees. We note that the model results are symmetric, therefore only the half width is plotted.} \label{fig:cme_interaction}
\end{figure}
Immediately after the disappearance of the second type-II burst at 03:12 UT, a broadband structure appears between 3:13 and 4:15\,UT with clearly drifting high- and low-frequency cutoffs (indicated by white dashed lines) as shown in the dynamic spectrum in Fig. \ref{fig:radio2}.
This whole structure, filled with type-III-like radio bursts, slowly drifts to lower frequencies with a drift rate comparable to the second type-II burst.
The drifting cutoffs are not harmonically related, suggesting that the low-frequency border could be a new propagating disturbance starting essentially higher in the corona than the second type-II burst and its driver.
As discussed in the following section this drifting type-II-like structure could be related to the interaction of the two associated \acp{cme}.
Later (see Fig. \ref{fig:radio}), after 4:15\,UT, the type-II burst extends far into the interplanetary medium until $>$14\,UT and frequencies $<$0.2\,MHz
and shows a clumpy broadband frequency structure implying an extended shock front occupying a broad range of densities during its propagation.
%
%
\subsubsection{The \ac{cme} interaction}\label{sec:cme_interaction}
Figure \ref{fig:gcs} shows the result of the Graduated Cylindrical Shell (GCS) model \citep{Thernisien2006, Thernisien2009}, applied to the multi-spacecraft coronagraph observations from \ac{stb} (left), SOHO (center), and \ac{sta} (right) between 3:15 and 3:21\,UT.
The top panel shows difference images and the bottom panel shows the same with the reconstructed \ac{cme} structures overplotted, where the green flux rope represents \ac{cme}1 and the orange one \ac{cme}2.
Based on the GCS reconstruction we derive that both \acp{cme} propagate in the same direction (\ac{cme}1: E140S25 and \ac{cme}2: E150S15 - uncertainties are within $\pm$10 degrees for longitude and latitude) which makes an interaction between their apexes highly likely. 
To derive the time and height of the apex interaction, we present in Fig. \ref{fig:cme_interaction} the reconstructed CME fronts as function of their angular width (nb: the tilt, which is off by 80 degrees between the two flux ropes, is not taken into account) and marked the interaction with a red flash.
From the model results we find that the apexes interact shortly after 3:15\,UT at a height of $d>$ 4\,R$_{\mbox{s}}$.
Although Fig. \ref{fig:cme_interaction} suggests an interaction of the flanks before 3:05\,UT at $d~\sim$~3\,R$_{\mbox{s}}$ we note that the tilt between the two flux ropes of about 80 degrees makes the flank interaction rather unlikely.
The radio observations, discussed in Sect. \ref{sec:remote}, show a drifting type-II-like structure with clear high- and low-frequency cutoffs containing type-III-like beams between 3:15 and 4:17\,UT. 
The heights of these cutoffs are 2.1 and 4.7\,R$_{\mbox{s}}$ \cite[using a 1x Saito model,][]{Saito1970} making it likely that these radio signatures are caused by the interaction of the CME apexes.
Furthermore, the type-III-like beams inside this drifting structure could be the signature of accelerated electrons between the two magnetic structures approaching each other.
The two type-II radio bursts during the beginning of the event ($t<$ 3:15\,UT) indicate the presence of two distinct shocks.
These shocks are likely associated to each of the two \acp{cme}.
We note that, while the flux ropes of the two \acp{cme} are not able to penetrate through each other, the \ac{cme}-driven shock of \ac{cme}2 might pass through the slower \ac{cme}1 \citep{Vandas1997}.
%
%
\subsection{In-situ observations}\label{sec:in-situ}
\subsubsection{Solar energetic particle observations}\label{sec:sep_obs}
\begin{figure*}[p] 
\centering
\includegraphics[width=0.45\textwidth, clip=true, trim = 0mm 20mm 0mm 30mm]{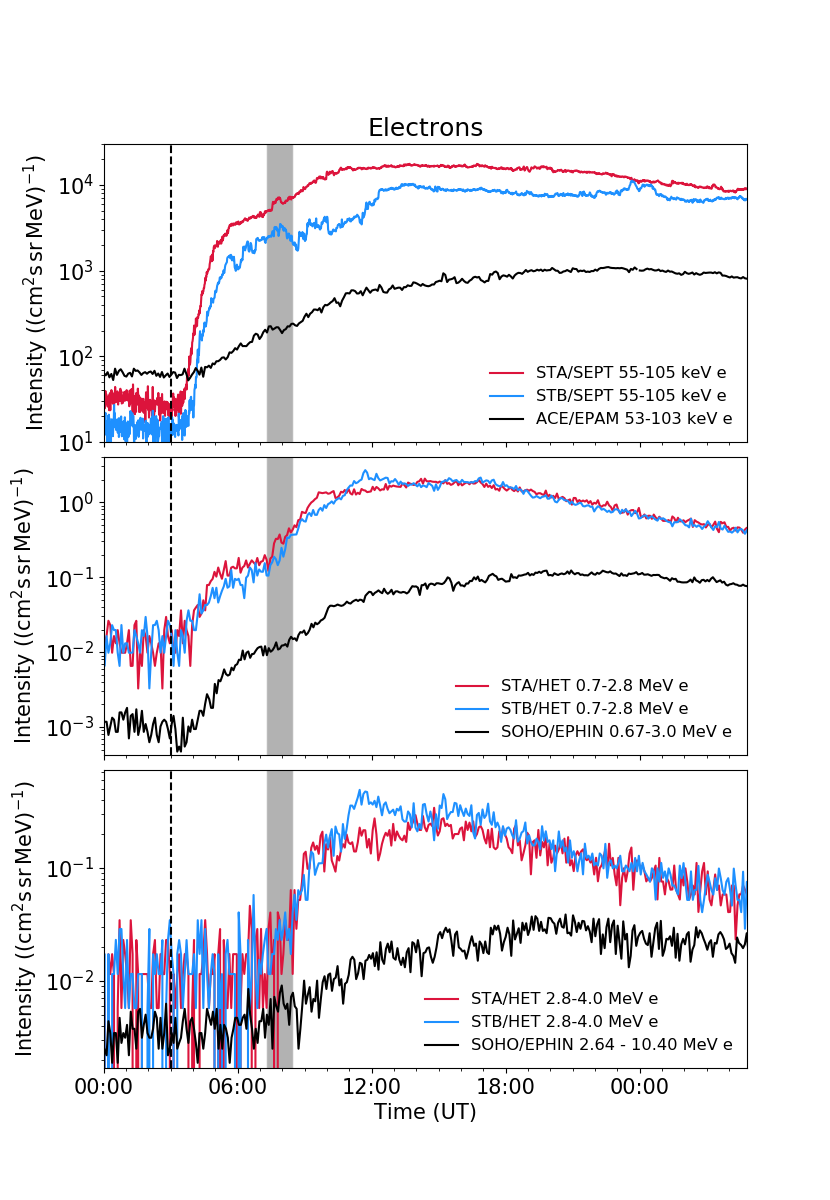}
\includegraphics[width=0.45\textwidth, clip=true, trim = 0mm 20mm 0mm 00mm]{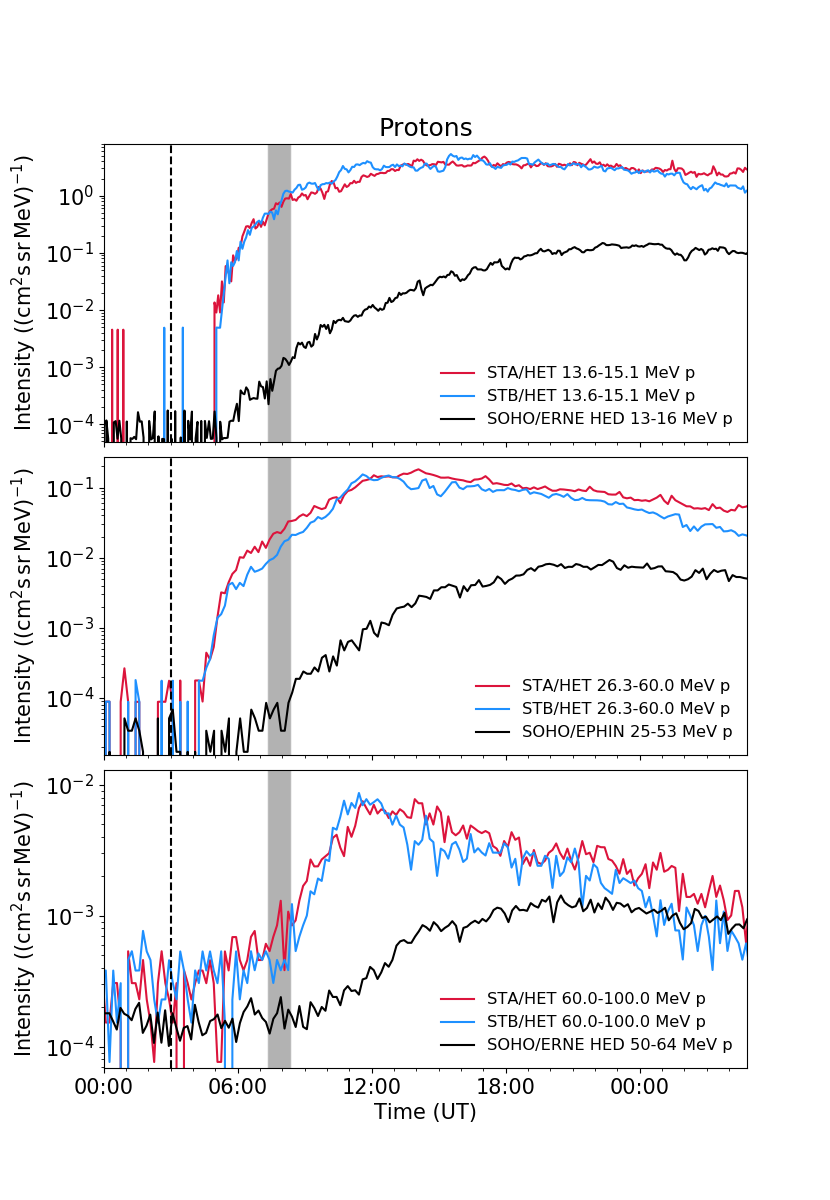}
 \caption{Solar energetic electron (left) and proton observations (right) of the 26 Dec 2013 \ac{sep} event. Each panel shows measurements at the two STEREO spacecraft and close to Earth at SOHO or ACE at comparable energy bins. Generally the energy increases from top to bottom. Intercalibration factors have been applied to the ACE/EPAM and SOHO/EPHIN electron measurements and to the EPHIN proton measurements following \protect\citet{Lario2013}. The shaded range marks the  time when the second component sets in (see text).} \label{fig:intro_particles}
\end{figure*}

The SEP observations at \ac{sta} (red), \ac{stb} (blue), and ACE/SOHO (black) on 26 Dec 2013 are shown in Fig. \ref{fig:intro_particles} for electrons (left panel) and protons (right panel). 
%
%
%
\begin{figure*}[p] 
        \includegraphics[width=0.349\textwidth, clip=true, trim = 0mm 15mm 10mm 0mm]{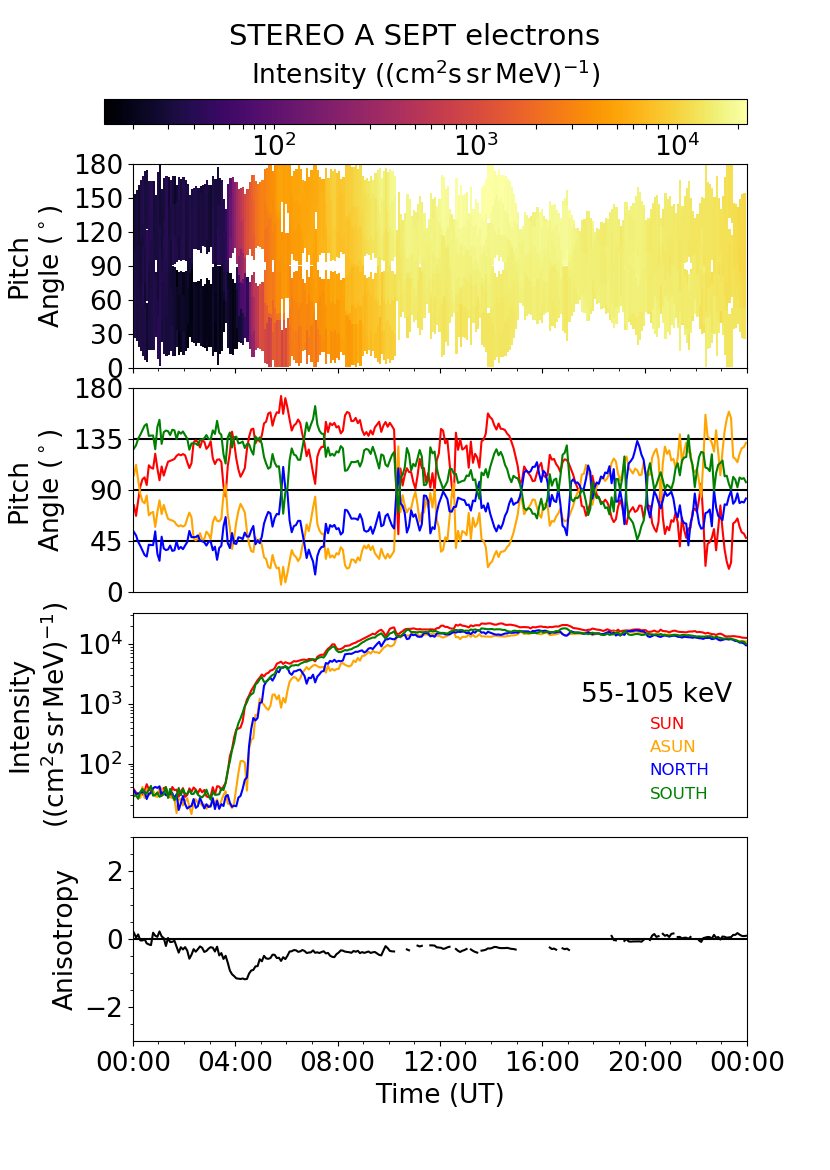}
        \includegraphics[width=0.315\textwidth, clip=true, trim = 19.9mm 15mm 10mm 0mm]{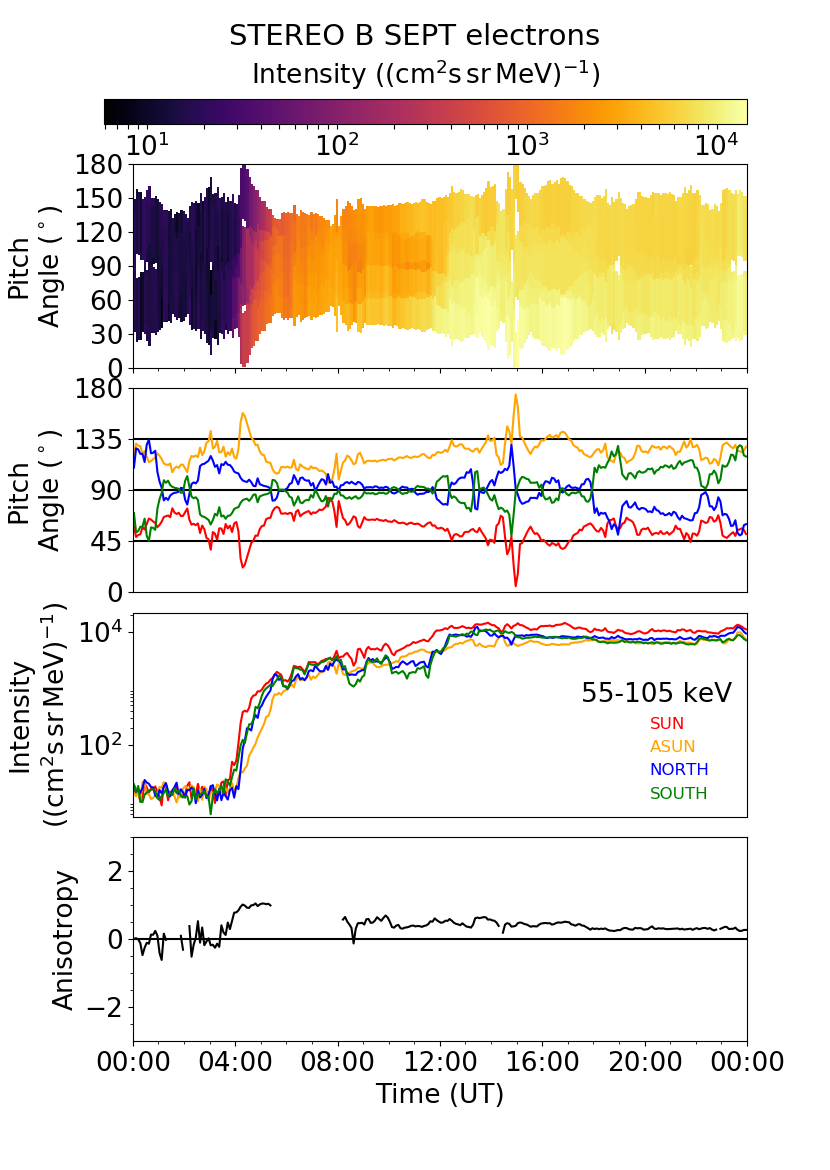}
        \includegraphics[width=0.315\textwidth, clip=true, trim = 19.9mm 15mm 10mm 0mm]{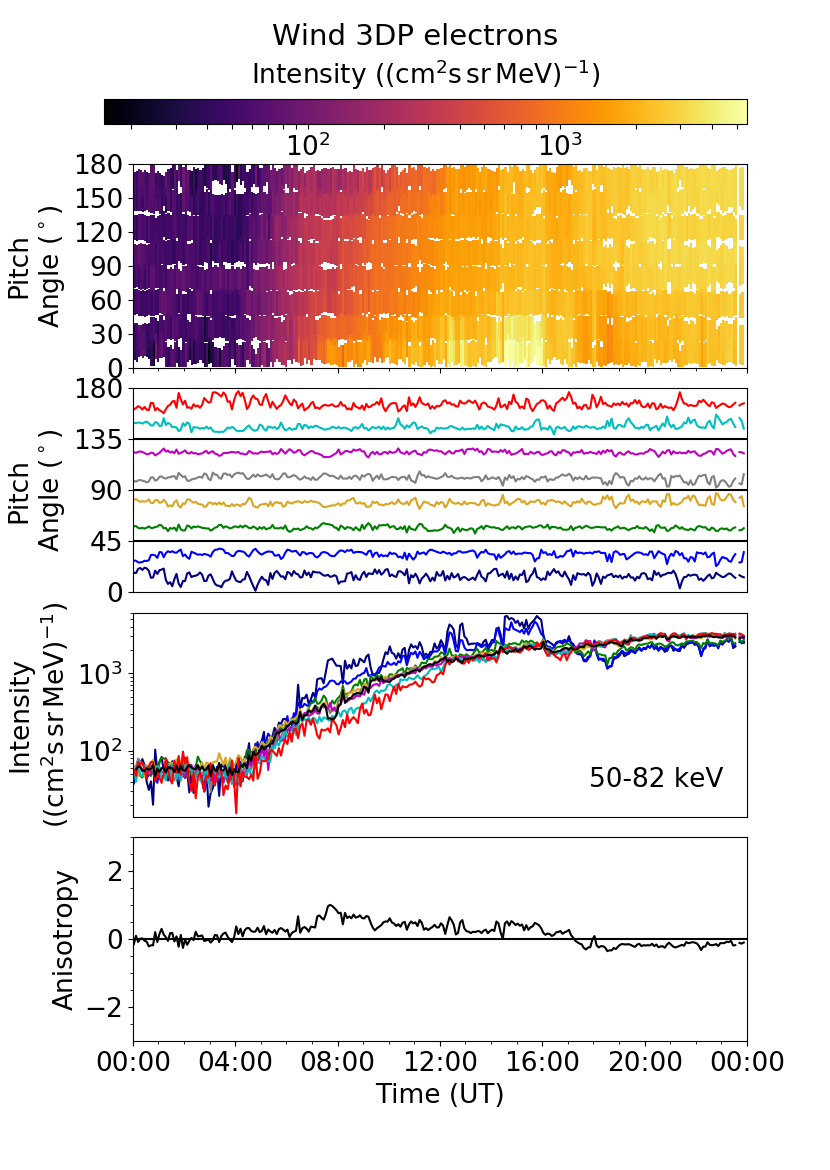}
        \caption{Sectored energetic electron observations at \ac{sta} (left), \ac{stb} (center), and Wind (right). Each plot shows from top to bottom the color coded pitch-angle-dependent intensity distribution, the pitch angles of each viewing sector provided by the instrument, the corresponding intensity of each viewing direction, and the first-order anisotropy index as determined from the data.} \label{fig:ani_plots} 
\end{figure*}%
From top to bottom the energy increases showing near-relativistic electrons (55-105\,keV) in the top left panel, and relativistic electrons (0.7-2,8\,MeV and 2,8-4 MeV) below.
We note that in the following we always use the terms `near-relativistic' and `relativistic' electrons for the 55-105\,keV and 0.7-2,8\,MeV energy channels, respectively.
The top right panel shows 13-16\,MeV protons, the middle panel $\sim$ 25-60\,MeV and the bottom right panel $>$60\,MeV and 50-60\,MeV for the two STEREO spacecraft and SOHO/ERNE, respectively.
We note that the ACE/EPAM and SOHO/EPHIN electron observations have been multiplied by intercalibration factors of 1/1.3 and 1/13. and EPHIN proton observations with a factor of 1.1 following \citet{Lario2013}.\\
Although the two STEREO spacecraft are separated by 59 degrees, the time series of the \ac{sep} intensities of electrons and protons are very similar at both spacecraft. 
Even at the position of Earth, longitudinally separated by 150 degrees to each of the STEREO spacecraft, clear energetic particle increases are observed up to relativistic energies ($>= 4$\,MeV for electrons).
However, as shown by the bottom right panel, no protons $>$60\,MeV are observed close to Earth while protons in the highest available energy channel of 60-100\,MeV are observed at both STEREO spacecraft.
The time profiles at all energies show long rise times of several hours up to almost a day.\\
The dashed black lines mark the onset of the associated type-III radio burst at 3:02\,UT.
The onset times of the near-relativistic electrons (top left panel) are 3:45\,UT (\ac{sta}), 3:53\,UT (\ac{stb}), and 4:10\,UT (ACE).
Assuming that the onset of the type-III burst marks the injection of the \acp{sep} at the Sun, these electrons arrive at the spacecraft with a delay of 23 (\ac{sta}), 31 (\ac{stb}), and 48 minutes (ACE), respectively, compared to the scatter-free transport along a nominal Parker spiral.\\
It is important to note, the onsets of the relativistic electrons are later than those of the near-relativistic electrons (see Table \ref{table:timeline}).
The highest-energy particles shown in the bottom panels arrive much later.
The near-relativistic electrons and low-energy protons show a more or less rapid rise followed by a more gradual increase. 
The higher the energy, the less prominent the first steep increase and the more prominent the second component which is accompanied by a new steepening of the intensity time series (see e.g., the middle panels).
The gray shaded range marks the approximate time of this steepening, indicating the onset of this second component, between 7:15\,UT and 8:30\,UT, that is, $\sim$4 hours later than the first component.
We note that the onset of this second component is uncertain (and might be even earlier) because of the first component masking the onset.
Interestingly, the first component nearly vanishes at 60-100\,MeV protons and 3-4\,MeV electrons (bottom panels) and only an increase corresponding to the second component is observed at these high energies (bottom panels). 
Even the energetic particle observations at the spacecraft close to Earth tend to show this break, suggesting the global character of this phenomenon which is observed all around the Sun.\\
Figure \ref{fig:ani_plots} displays the sectored intensity measurements of near-relativistic electrons as observed by \ac{sta} (left), \ac{stb} (center), and Wind (right). 
The top panel of each plot shows the pitch-angle-dependent intensity distribution where the color-coding corresponds to the electron intensity. 
White areas denote pitch angle ranges which were not covered by the telescopes taking into account the opening angles of the apertures. 
The second panel shows the pitch angles of the centers of the available viewing directions, the third panel displays the corresponding intensity measured in those viewing directions, and the bottom panel shows the first order anisotropy index as computed from the above observations \citep[see e.g.,][]{Dresing2014}.
Figure \ref{fig:ani_plots} clearly shows that all three spacecraft observed significant anisotropies. 
At the two STEREO spacecraft the anisotropy is larger during the first few hours of the event and then reduces to a lower level.
At Wind, however, the pitch angle distribution is rather isotropic during the first phase of the event but becomes more anisotropic around 7\,UT. 
The reason for this could either be different propagation conditions during the two phases or a change of the source size, meaning that it gets closer to the magnetic field line connecting to Wind. 
We note, however, that significant anisotropy can be even observed without a direct magnetic connection to the source region close to the Sun \citep[e.g.,][]{Strauss2017}.\\
Figure \ref{fig:ani_plots} also shows that the total anisotropic period in the near-relativistic electron event extends over many hours at all three viewpoints.
\ac{sta} observes at least 14 hours of anisotropic flux. 
Unfortunately, after this time ion contamination begins to alter the electron measurement, so that it is no longer possible to determine a reliable anisotropy. 
For \ac{stb} the ion contamination sets in much later so that an anisotropy lasting for a period of over a day can be confirmed. 
Even at Wind, which is situated on the far side of the associated activity region at the Sun (cf. Fig. \ref{fig:constellation}), significant anisotropy is observed over nearly 12 hours. 
\subsubsection{Interplanetary context}\label{sec:IPcontext}
\afterpage{\clearpage} 
\begin{figure*}[p] 
	\begin{adjustbox}{addcode={\begin{minipage}{\width}}{
\caption{%
      Solar wind plasma and magnetic field observations at \ac{stb} (a), \ac{sta} (b), and ACE (c) from 25 Dec 2013 until 31 Dec 2013. From top to bottom: $<$100\,keV electron intensities (gray shades mark periods of ion contamination), latitudinal, and azimuthal angles of the magnetic field, magnetic field strength, proton temperature, proton density, and solar-wind speed. The colored band below each plot represents the magnetic field polarity with red (green) showing inward (outward) polarity and yellow standing for uncertain polarity periods. The passages of \ac{icme}s and shocks are marked by blue shaded ranges and lines (see text).}\label{fig:overview_STB_STA}
  \end{minipage}},rotate=90,center}
\subfloat[]{\includegraphics[width=0.55\textwidth, clip=true, trim = 0mm 0mm 0mm 0mm]{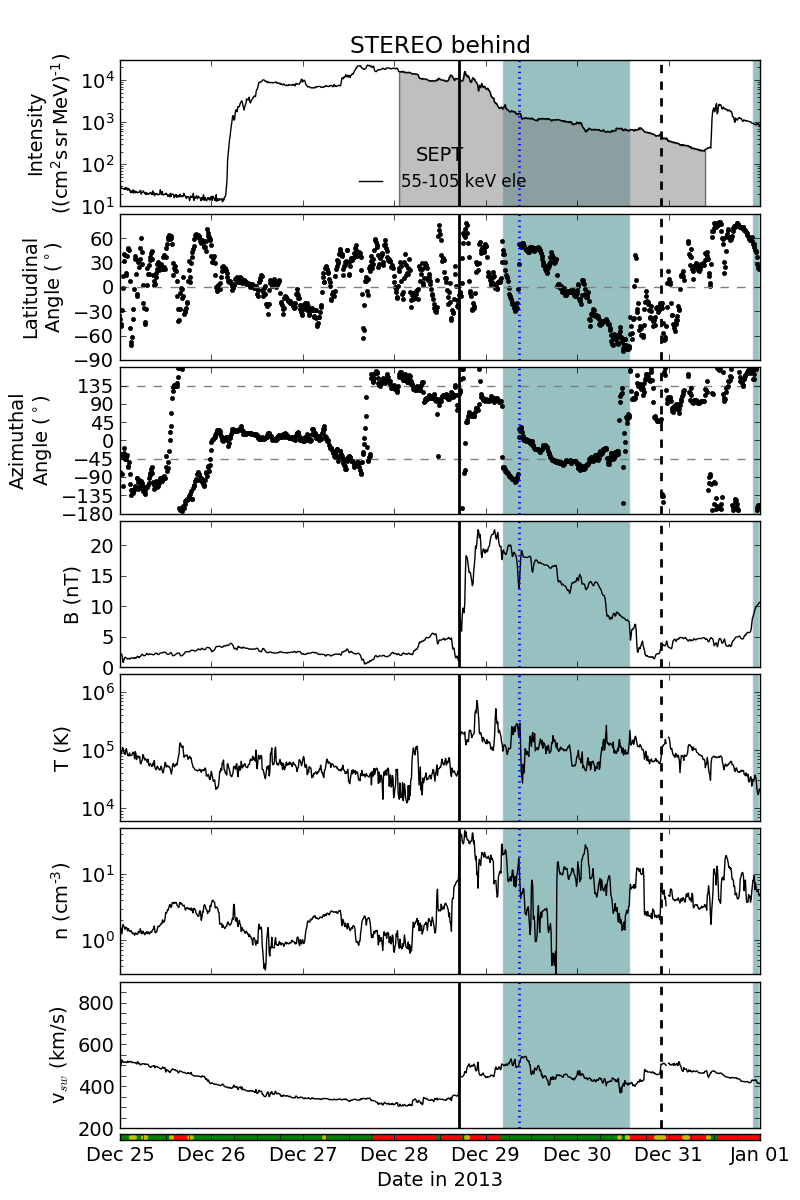}}
\subfloat[]{\includegraphics[width=0.47\textwidth, clip=true, trim = 30mm 0mm 0mm 0mm]{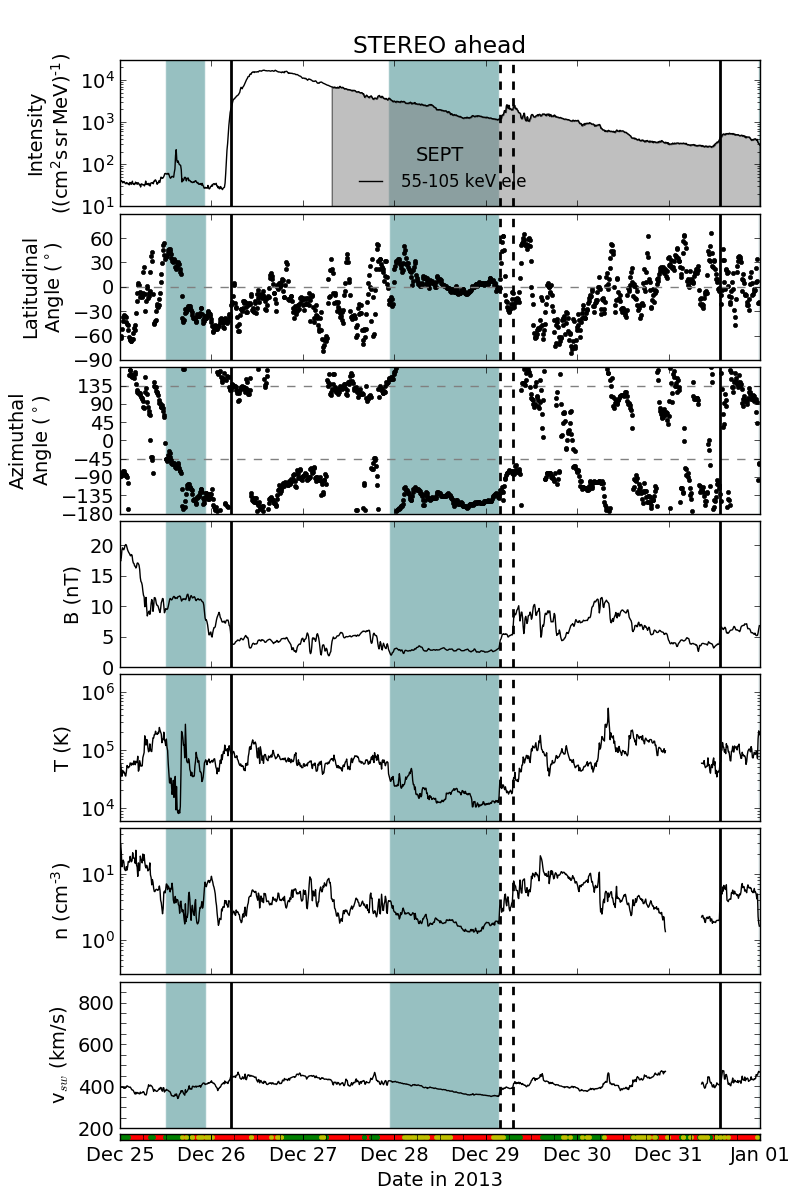}}
\subfloat[]{\includegraphics[width=0.47\textwidth, clip=true, trim = 30mm 0mm 0mm 0mm]{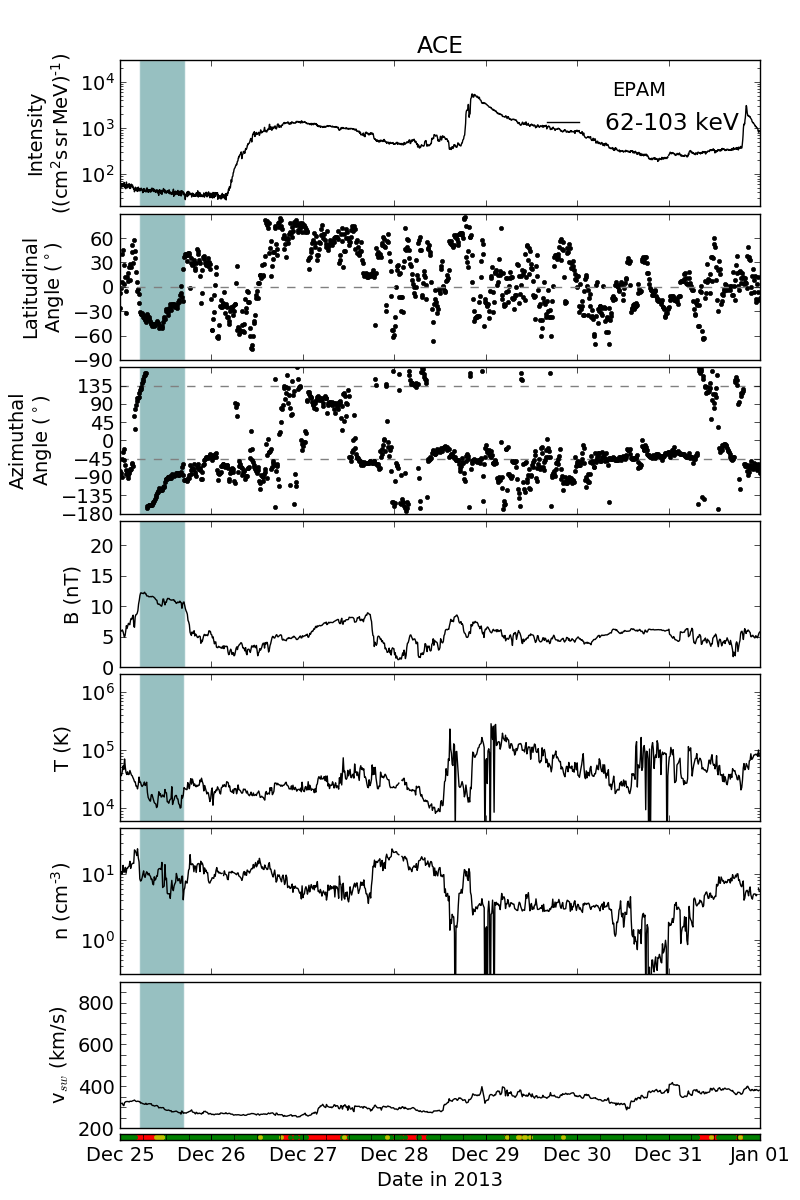}}
    \end{adjustbox}
\end{figure*}%
%
Figure \ref{fig:overview_STB_STA} shows solar-wind plasma and magnetic field observations during the period of Dec 25 to Dec 31 2013 for \ac{stb} (a), \ac{sta} (b) and ACE (c).
To guide the eye, each top panel shows the electron event in the 55-105\,keV SEPT and 62-103\,keV EPAM channels.
Periods where the electron measurements are contaminated by ions (and therefore cannot be trusted) are marked with gray shades.
The two panels below display the magnetic field longitudinal and azimuthal angles in the RTN coordinate system. 
Below the magnetic field magnitude, the solar wind proton temperature and density and the solar-wind speed are presented.
The colored band below each plot marks the in-situ magnetic field polarity with red meaning inward magnetic field, green outward field and yellow standing for periods where a unique assignment is not possible.
While \ac{stb} observes the event in a positive polarity sector, \ac{sta} is situated in a negative sector. 
This also causes the different pitch angle ranges where the first arriving particles are observed at \ac{sta} (pitch angle 180) and \ac{stb} (pitch angle 0); see Fig. \ref{fig:ani_plots}.
At the very beginning of the event, ACE also observes positive polarity but a rotation of the magnetic field follows, leading to unclear polarity regions and later to negative polarity later in the day of 27 Dec. \\
In all plots of Fig. \ref{fig:overview_STB_STA} the passage of shocks and interplanetary coronal mass ejections (\ac{icme}s) are marked by vertical black lines and shaded ranges, respectively.
The times were provided by the STEREO \ac{icme} and shock catalogs by Jian et al.\footnote{\url{http://www-ssc.igpp.ucla.edu/forms/stereo/stereo_level_3.html}} and the Near-Earth \ac{icme} list by Richardson and Cane\footnote{\url{http://www.srl.caltech.edu/ACE/ASC/DATA/level3/icmetable2.htm}}.\\
The onset of the \ac{sep} event at \ac{stb} happens during a rather quiet period of decreasing solar wind speed, low magnetic-field strength, and weak magnetic-field variations.
Indeed a magnetic-field rotation is visible, suggesting the presence of a magnetic flux tube not listed in the above catalogs.
From 29 Dec 4:12\,UT to 30 Dec 14:00\,UT \ac{stb} observes an \ac{icme} (shaded range) which drives a shock passing the spacecraft at 17:06\,UT on 28 Dec (solid line).
This \ac{icme} is likely the one associated to the analyzed event (\ac{cme}2). 
Indeed two distinct magnetic flux ropes can be identified showing different orientations in the magnetic field angles likely associated to \ac{cme}1 and \ac{cme}2.
The border between these two \acp{icme} is marked by a blue dotted vertical line (see Fig. \ref{fig:overview_STB_STA} a).
The later shock on 30 Dec (dashed line) is attributed to fast solar wind overtaking an I\ac{cme} (according to the shock catalog by Jian et al.\footnote{\url{http://www-ssc.igpp.ucla.edu/forms/stereo/stereo_level_3.html}})
\\
The interplanetary context observed at \ac{sta} is more complex.
The shaded range starting on Dec 25 marks an \ac{icme} embedded in a \ac{sir}.
Shortly after the \ac{sep} onset the \ac{sir}-associated reverse shock is observed at 05:04\,UT.
The later double vertical line on 29 Dec marks again two \ac{sir}-associated forward shocks.
The latest vertical line marks a \ac{cme}-driven shock.
The shaded range on Dec 28 marks the transit of an \ac{icme} not accompanied by a shock.
This \ac{icme}, passing \ac{sta} from 27 Dec 22:22\,UT to 29 Dec 3:38\,UT, is likely the one associated to the analyzed event.
However, compared to \ac{stb}, the magnetic cloud signatures are less pronounced and no distinct flux ropes can be identified.
This is in agreement with the propagation direction of \ac{cme}1 and \ac{cme}2 towards \ac{stb} hitting \ac{sta} with its flank.\\
During the whole period shown in Fig. \ref{fig:overview_STB_STA} (c), ACE observes slow solar wind.
No shocks were observed at any of the close to Earth spacecraft but an \ac{icme} was observed at ACE on 25 Dec, well before the onset of the gradual increase of the 26 Dec \ac{sep} event. 
In contrast to the STEREO spacecraft ACE also observes impulsive \ac{sep} events on 24 Dec and on 28/29 Dec.
%
\begin{figure*}[h] 
        \centering
        \includegraphics[width=0.98\textwidth, clip=true, trim = 0mm 0mm 0mm 0mm]{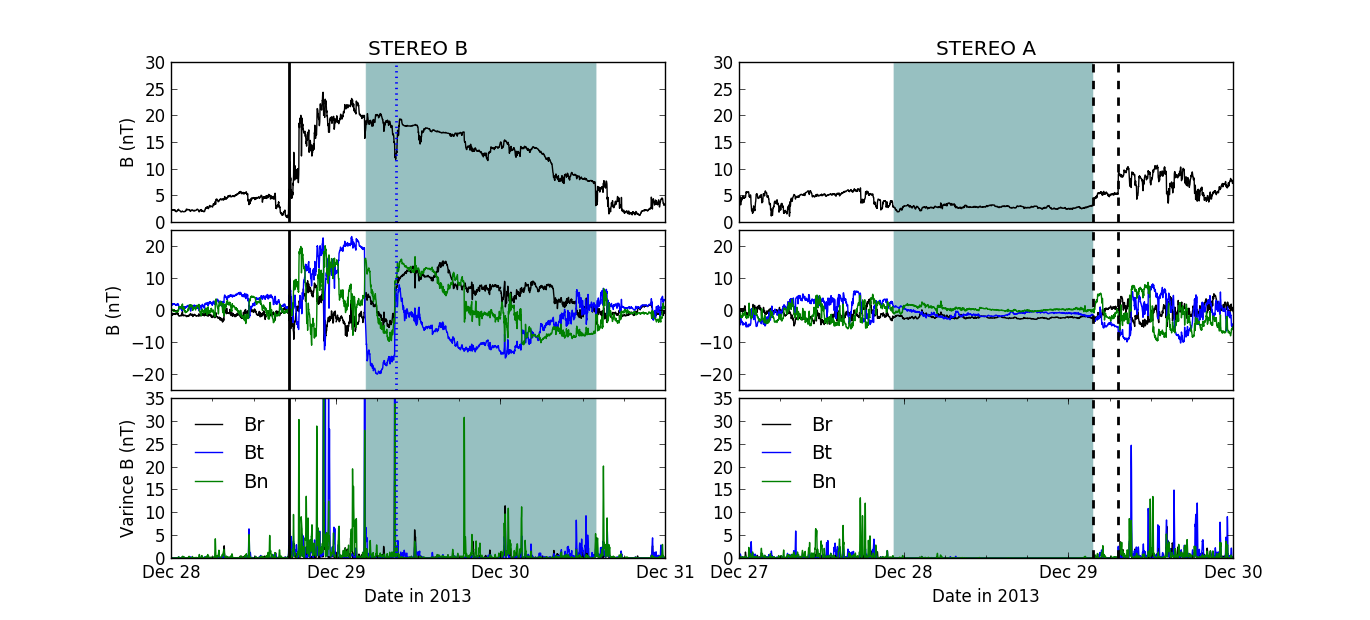}
        \caption{Magnetic field observations at \ac{stb} (left) and \ac{sta} (right). From top to bottom: Magnetic field magnitude, RTN components, and the variances of the components computed using a sliding window of 10 minutes. Shades and lines are as in Fig. \ref{fig:overview_STB_STA}.} \label{fig:var}
\end{figure*}%
\\
Figure \ref{fig:var} shows the magnetic field magnitude (top) and RTN components (middle) at \ac{sta} and \ac{stb} during the passage of the associated \ac{icme}. While the magnetic field variations inside the \ac{icme} observed at \ac{sta} are very low, an unusual high variance is observed at \ac{stb}.
This might have been caused by the interaction process of the two \ac{cme}s.
The variances of the RTN components (also Fig.11) could be the signature of steady reconnections that occurred during its outward propagation which would cause the smooth magnetic field rotation normally observed inside magnetic clouds to  vanish.
%
%
%
\subsection{The longitudinal spread of the \ac{sep} event}\label{sec:spread}
\begin{figure*}[ht] 
\begin{minipage}{.35\linewidth}
\centering
\subfloat[]{\includegraphics[width=\textwidth, clip=true, trim = 0mm 0mm 0mm 0mm]{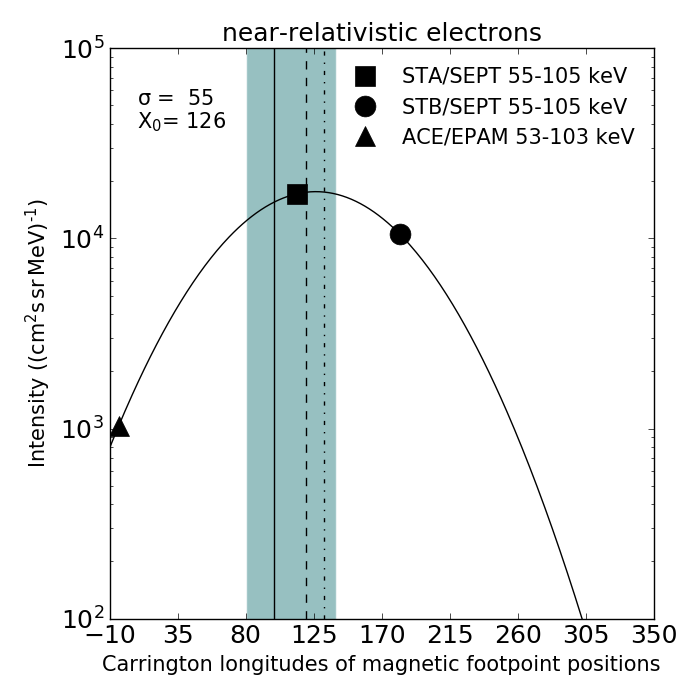}}
\end{minipage}%
\begin{minipage}{.32\linewidth}
\centering
\subfloat[]{\includegraphics[width=\textwidth, clip=true, trim = 15mm 0mm 0mm 0mm]{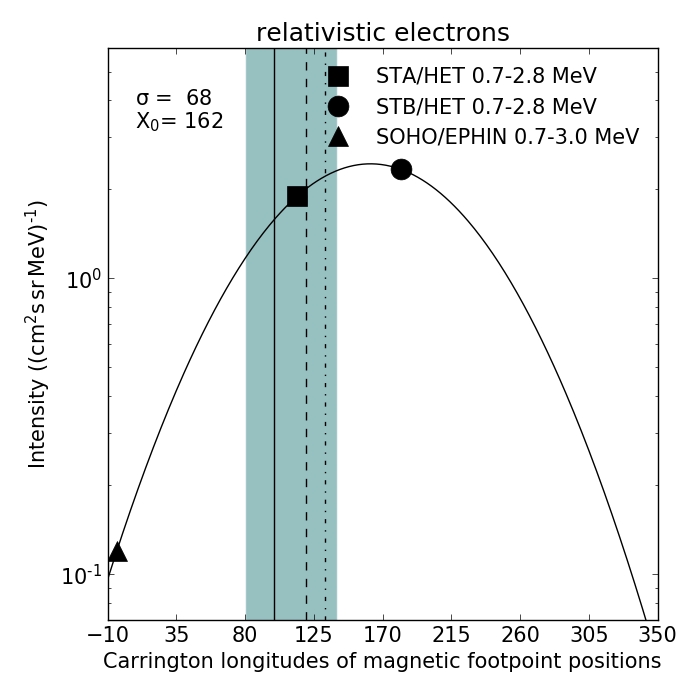}}
 \end{minipage}
\begin{minipage}{.32\linewidth}
\centering
\subfloat[]{\includegraphics[width=\textwidth, clip=true, trim = 15mm 0mm 0mm 0mm]{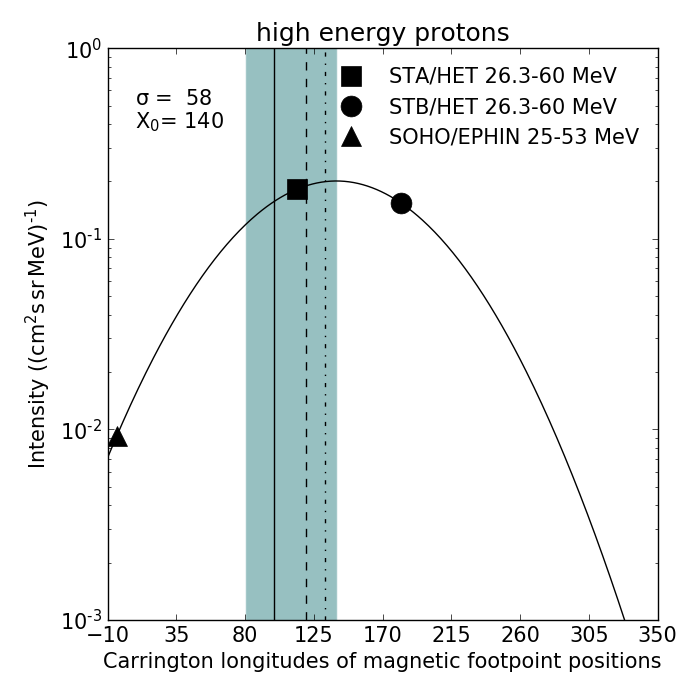}}
 \end{minipage}
 \caption{Longitudinal distribution of peak intensities of near-relativistic electrons (a), relativistic electrons (b), and $>$25\,MeV protons (c) observed by STA (squares), \ac{stb} (circles), and SOHO/ACE (triangles). The peak intensities are plotted as a function of the magnetic footpoint positions in Carrington longitudes determined by ballistic backmapping. The point of SOHO/ACE is plotted at -4 degrees instead of 356 degrees Carrington longitude to retain the correct ordering of the observers. A Gaussian function has been applied to the data points with its standard deviation $\sigma$ and its center $X_0$ provided in the top-left corners of each plot. The shaded range marks the longitudinal extent of the whole complex coronal event at the Sun. The solid vertical line marks the longitude of the main flare in region \#2, the dashed line represents the longitude of the associated \ac{cme}, and the dash-dotted line marks the longitude of the activity region \#1.} \label{fig:gaussians} 
\end{figure*}
\begin{figure}[b!] 
        \includegraphics[width=0.5\textwidth, clip=true, trim = 0mm 0mm 0mm 0mm]{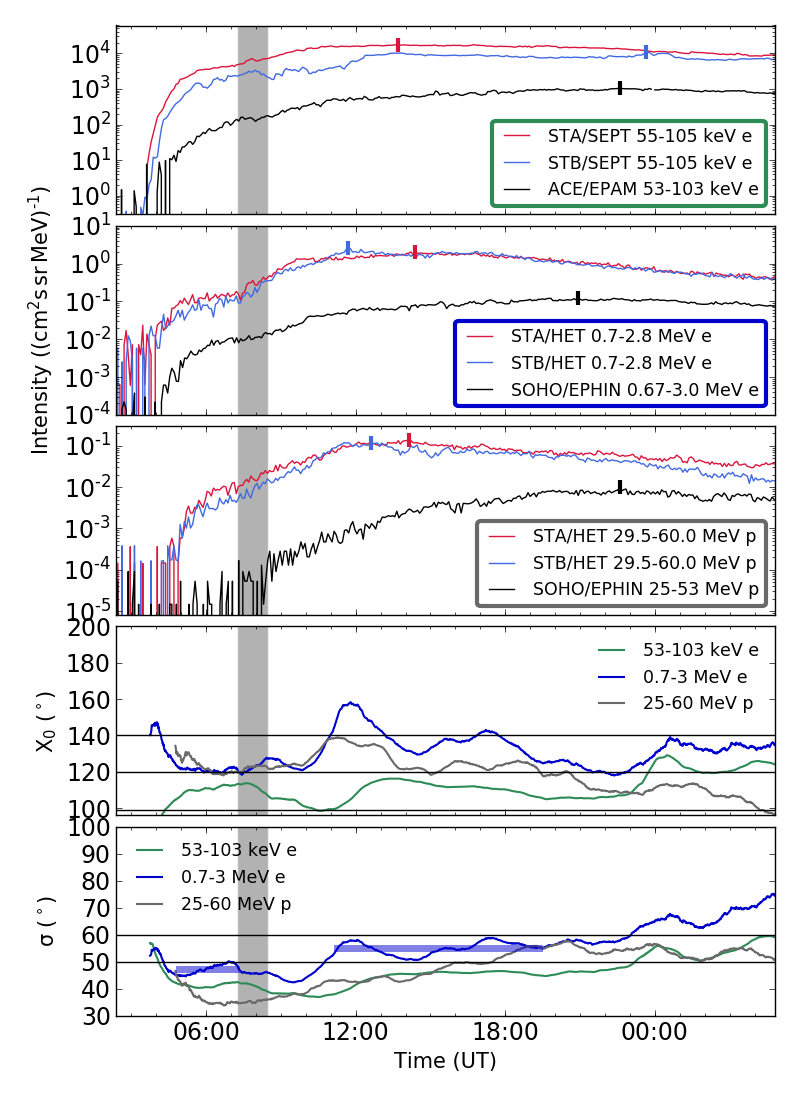}
        \caption{From top to bottom: The first three panels display the intensity time series of near-relativistic electrons, relativistic electrons, and $\sim$30-60\,MeV protons observed at STA (red), \ac{stb} (blue), and SOHO or ACE (black) on 26 Dec 2013. The two panels below show the center $X_0$ and the standard deviation $\sigma$ of a Gaussian function applied to the intensity measurements at each time step using a one-hour averaging window. The small vertical dashes in the top three panels represent the peak intensities taken to produce Fig. \ref{fig:gaussians}. The light blue bars in the bottom panel denote the mean standard deviations of the relativistic electron distributions during the first and second components.} \label{fig:running_gauss} 
\end{figure}

%
Owing to the observations at the three well-separated viewpoints, the 26 Dec 2013 event can unambiguously be proven as a widespread event. 
Energetic electrons and protons were observed well above background over a longitudinal range of 210 degrees (spanned by \ac{stb} over STA to the position of Earth).
To characterize the width of an \ac{sep} event it has become common to apply a Gaussian function to the observed peak intensities at the different viewpoints as a function of the longitudinal separation angle \citep[e.g.,][]{Lario2006, Lario2013, Richardson2014, Dresing2014}, being the angle between the spacecraft magnetic footpoint at the Sun and the longitude of the source region, that is, the flare. 
The standard deviation of these Gaussians usually ranges between $\sim$30 and 50 degrees for widespread electron events \citep{Lario2013, Dresing2014, Richardson2014, Gomez-Herrero2015}.
The event from 25 Feb 2014 was one of the widest events in that sense with a sigma of 57 degrees (for 71-112\,keV electrons, \citet{Lario2016}) or 47 degrees for 0.7-3\,MeV electrons assuming a symmetric Gaussian distribution \citep{Klassen2016}.
In our case it is difficult to identify a precise source region because of the large area involved at the Sun during the complex coronal event.
We therefore plot in Fig. \ref{fig:gaussians} the observed peak intensities as a function of the longitude (in Carrington coordinates) of the spacecraft magnetic footpoints at the Sun (instead of the separation angle). 
We note that the pre-event background has been subtracted and an intercalibration between the STEREO spacecraft and SOHO/ACE has been applied following \citet{Lario2013}.
The standard deviations of the applied Gaussians are very large with 55 degrees for near-relativistic electrons (Fig. \ref{fig:gaussians} (a)), 68 degrees for relativistic electrons (Fig. \ref{fig:gaussians} (b)) and 58 degrees for 25-60\,MeV protons (Fig. \ref{fig:gaussians} (c)) making this alongside the 25 Feb 2014 event the widest \ac{sep} event ever observed.
Because the event is observed well above background at all three positions, the Gaussian functions displayed in Fig. \ref{fig:gaussians} suggest that the actual event may have extended all around the Sun at 1\,AU, like, for example, the circumsolar event reported by \citet{Gomez-Herrero2015}.\\
Figure \ref{fig:gaussians} also shows that the center of the distribution $X_0$ is not at the same longitude for near-relativistic and relativistic electrons.
The distribution of the relativistic electrons (b) is shifted towards west, centering at Carrington longitude 162, that is, a shift of 36 degrees with respect to the near-relativistic electrons (a)).
The $>$25\,MeV proton distribution also shows a shift towards west with a center at 140 degrees (c).
The shaded range in the plots marks the extent of the associated complex coronal event, that is, the assumed source region of the \acp{sep}.
The solid black line marks the longitude of the main flare of the event at 99 degrees (region \#2, see Sect. \ref{sec:remote}), and the dashed line represents the longitude of the associated \ac{cme} (\ac{cme}2) which is directed towards \ac{stb} at 120 degrees Carrington longitude.
The dashed-dotted line marks the longitude of the \ac{ar} associated to the preceding \ac{cme} (\ac{cme}1, region \#1) at 132 degrees.
While the center of the distribution of the near-relativistic electrons fits within the longitudes of the two \acp{cme} and the range of the Sun involved in the coronal event, the relativistic electrons and protons have their centers outside the shaded range, westward of it.
We note that none of the distributions is centered around the longitude of the main flare (solid line).
An asymmetry towards western longitudes is predicted by transport modeling \citep{He2011, Strauss2015} due to perpendicular diffusion effects  with a growing shift with stronger perpendicular diffusion \citep{Strauss2017}.
The statistical analysis of multi-spacecraft \ac{sep} observations by \citet{Lario2006, Lario2013} also suggested a shift of the peak-intensity distribution to the west of the order of 10 to 15 degrees.
The latter authors attributed this asymmetry to the presence of the \ac{cme} shifting the spacecraft magnetic connections to stronger portions of the shock.
However, similar statistical studies \citep[e.g.,][]{Dresing2014} found shifts towards east or suggested that a significant asymmetry cannot be determined because of the large variety from event to event \citep{Richardson2014}.
The large westward shifts in the present analysis, however, together with the strong anisotropies suggest that the observed asymmetries are not mainly caused by perpendicular diffusion effects but rather by a shift of the source location.\\
However, one must bear in mind that the assumption of the longitudinal \ac{sep} distribution following a Gaussian form (which we apply here) may be only partly correct \citep[e.g.,][]{Klassen2016}.
Furthermore, the peak intensities used for the Gaussian approximation may have been measured at different times meaning that the displayed distribution does not represent the real \ac{sep} distribution in space, and temporal changes are not resolvable.
We therefore analyzed the temporal evolution of the \ac{sep} distributions in Fig. \ref{fig:running_gauss} by applying a Gaussian function to each of the omni-directional intensity values measured by the three observers at the same times. 
The spacecraft motion with time and the measured solar-wind speed have been taken into account to determine the magnetic footpoint positions at the Sun allowing for a more accurate determination of the Gaussian distributions.
%
The figure shows in the top three panels the intensity time series at the three observers for the same particle and energy combinations like in Figure \ref{fig:gaussians} with STA in red, \ac{stb} in blue, and SOHO/ACE in black.
The small vertical bars denote the peak intensities used to determine the Gaussian distributions shown in Fig. \ref{fig:gaussians}. We note the strongly differing times of the occurrence of the peak intensities.
The two bottom panels of Fig. \ref{fig:running_gauss} present the parameters of Gaussian distributions applied to each time step using a running window of one hour: The center of the Gaussian distributions $X_0$ (second bottom panel) and its standard deviation $\sigma$ (bottom panel) for near-relativistic electrons (green), relativistic electrons (blue), and protons (gray).
The bottom panel shows that the longitudinal distributions were large throughout the whole event, all being $\sigma>=35$ degrees. 
A steady increase of the standard deviations throughout the event can be observed. 
We note that such broadenings in the late phase of \ac{sep} events are attributed to the so called reservoir effect \citep{McKibben2005, Zhang2003}.
The centers of the distributions shown in the panel above show significant variation with time. 
The gray bar represents the onset of the second component. 
Although no sudden change in $X_0$ and $\sigma$ can be observed directly after the gray bar, $\sim$ two hours later a shift of $X_0$ towards larger Carrington longitudes (towards west) and an increase of $\sigma$ are seen in the relativistic electron distribution (blue trace). 
The two-hour delay might be caused by the fact that the distribution of the first component is mixing with that of the second component especially during the early phase of the second component. 
Later, the second component dominates the first one causing the changes in the distributions to become visible. 
The two blue horizontal bars in the bottom panel (see figure caption for details) denote a clear increase of the mean standard deviation from the first ($\sim$47) to the second (55) component, respectively.
The distribution of the protons (gray traces) shows a similar behavior in $X_0$ although the standard deviation is significantly lower. 
The center of the distribution of the near-relativistic electrons (green traces) deviates from the other two distributions with more eastern longitudes.
The observations therefore show that during the same \ac{sep} event energetic particles of different energies and species may be distributed differently in space.
We note that different energy spectra at the different observers would also result in this effect.
Moreover, in the analyzed event, the source region and injection history of the near-relativistic electrons is likely not the same as the ones of the higher-energy particles.
The strong variations of the Gaussian parameters may be caused on the one hand by spatial and temporal variations of the injection function but transport effects could also play a role.
We note again that a Gaussian function may not perfectly represent the real particle distribution in space, inducing an uncertainty to the values shown in Fig. \ref{fig:running_gauss}.
However, short-time fluctuations are eliminated due to the one-hour averaging window and therefore the temporal evolution and long-term changes of the distributions should be represented reasonably.
%
%
%
%
%
\section{Discussion: The source of the \ac{sep} event}\label{sec:second_comp}
The long-lasting near-relativistic electron anisotropies and rise times of the 26 Dec 2013 event are unusual and require a prolonged electron injection.
Furthermore, the small onset delays between the three different spacecraft and their similar time profiles (e.g., at 0.7-3\,MeV electrons; see Fig. \ref{fig:intro_particles}) and especially the strongly similar intensity time series at the two STEREO spacecraft at many different energies (longitudinally separated by 59 degrees) suggest an extended injection region or an unusually quick particle spread close to the Sun.
A long-lasting injection could either be provided by an ongoing acceleration process, for example, by an extended shock front propagating into the IP medium, or by a leakage process from a large particle trap.\\
The second component observed in the \ac{sep} intensity time profiles about four hours later than the first component (cf. Fig. \ref{fig:intro_particles}) requires either a new \ac{sep} injection or a strong increase of the efficiency of the accelerator or the leakage process.
However, the higher the particle energy, the more prominent the increase of the second component.
The highest-energy particles, that is, $\sim3$\,MeV electrons and 60-100\,MeV protons were not even observed above background during the first component but only during the second one (see Fig. \ref{fig:intro_particles}). 
This fact suggests that the acceleration and injection mechanisms of the two \ac{sep} components may be different.
A pure strengthening of the original accelerator or stronger leakage are therefore rather unlikely.
This also means that the \ac{cme}-driven shock alone is not able to explain all the observations including the second component \acp{sep}.\\
We therefore assume that the second component is caused by a new distinct injection, different to the one responsible for the first component.
To identify the source of this later injection we survey the solar phenomena prior to the occurrence of the second component.
Although the associated solar event is complex, involving several regions of activity at the Sun (see Fig. \ref{fig:constellation_EUV} and Sect. \ref{sec:remote}), the flaring activity stops well before the appearance of the second component: the flare in region \#4 lasts only from 4:00 to 4:30\,UT and the  thermal EUV emission from the cooling (post-) flare loop arcade of the main two-ribbon flare reaches its maximum at 4:00 UT, and has substantially decayed around 4:45\,UT.
We note that a scatter-free path along a nominal Parker field line would correspond to a propagation time of $\sim$20 minutes for 55-105\,keV electrons and only of $\sim$10 minutes for 0.7-3.0\,MeV electrons.
Even if strong scattering would prolong the propagation time, the delay of three to four hours until the second component is observed is too large to be associated with the initial solar activity.
We find only one further potential candidate:
At 6:55\,UT a C2.2 flare is observed at W28 as seen from Earth (not shown here). 
It is associated to a series of type-III radio bursts occurring from 6:45 to 7:00\,UT observed by Wind/WAVES (see Fig. \ref{fig:radio}) and only very weakly observed at the STEREO spacecraft.
A \ac{cme} (\ac{cme}3 in Table \ref{table:timeline}) with a width of 129 degrees and a linear speed\footnote{\url{https://cdaw.gsfc.nasa.gov/CME_list/UNIVERSAL/2013_12/univ2013_12.html}} of 458\,km/s is associated to this event appearing at 7:24\,UT in LASCO/C2 and already at 7:05\,UT in the STA/COR1 field of view.
Because of its low speed it likely does not drive a shock which is in agreement with the absence of an associated type-II radio burst.
Although the timing of this activity roughly fits with the onset of the second component, we can exclude this flare as the parent source of the later \ac{sep} injection for the following reason:
The relative intensity ratios between the STEREO spacecraft (observing a more intense event) and at L1 spacecraft are the same during the first and the second component. 
If the source region of the second component were at the location of the small flare described above, that is, at the visible disk as seen from Earth, we would expect to observe a stronger intensity increase at L1 compared to the STEREO spacecraft.
We therefore suggest that the source region of the second component has to be close to the one of the first component (this is also reflected by Figs. \ref{fig:gaussians} and \ref{fig:running_gauss}). 
Because it is observed globally at all three viewpoints around the Sun it is likely that the parent injection region is large, at least covering the longitudinal separation between the two STEREO spacecraft producing nearly the same \ac{sep} intensity time profiles at that spacecraft.
We note that \citet{Kocharov2017} also suggested a trapping scenario to be responsible for the high-energy component of two \ac{gle} events.
However, the opening of the trap happened much earlier in these cases.\\
%
%
\begin{figure}[h!] 
        \includegraphics[width=0.45\textwidth, clip=true, trim = 0mm 45mm 0mm 0mm]{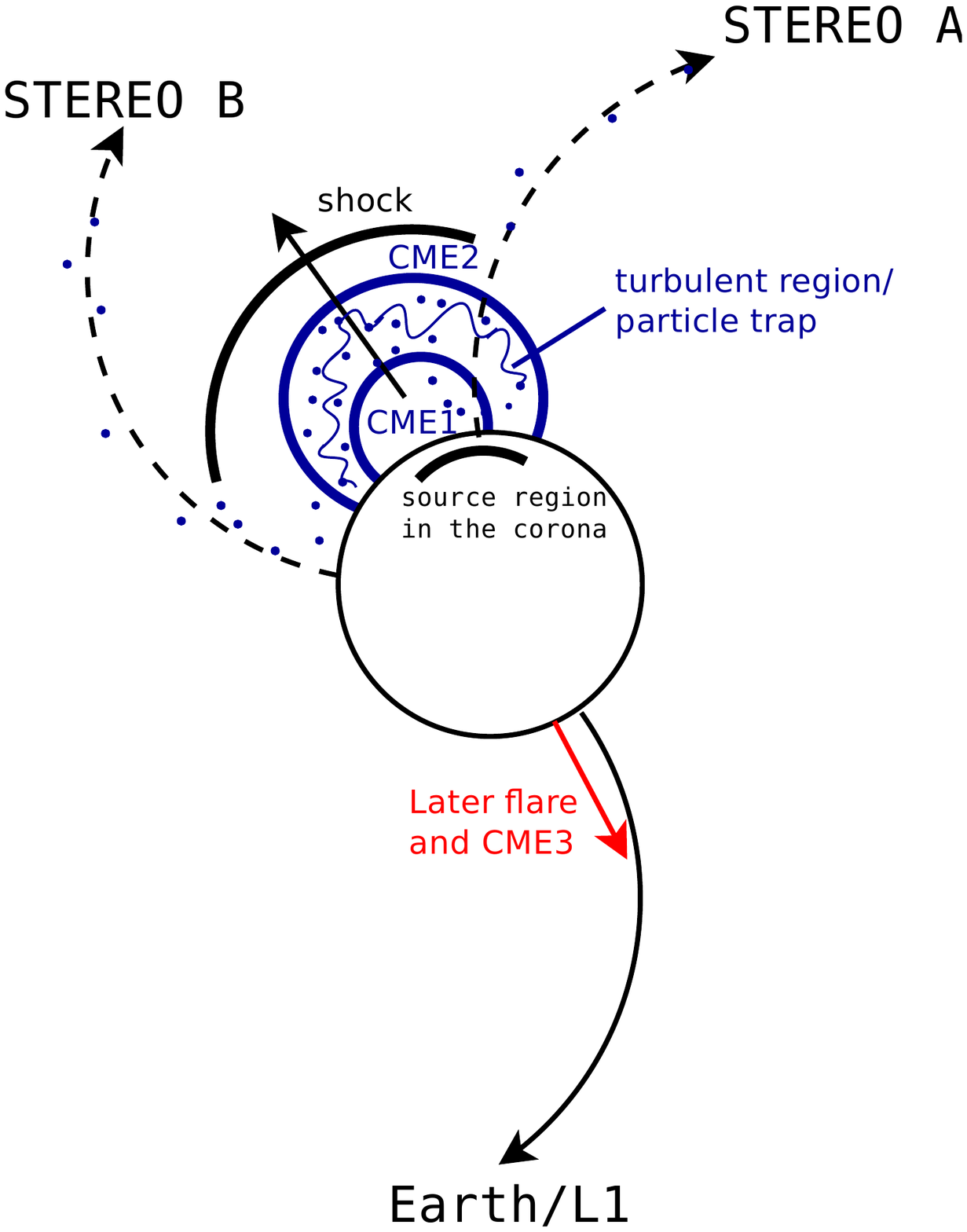}
        \caption{Cartoon illustrating the proposed scenario showing the longitudinal constellation of the two STEREO spacecraft and L1 with the turbulent region between \ac{cme}1 and \ac{cme}2 (blue) serving as a particle trap with steady particle leakage. The shock has already propagated through that region and is now driven in front of the merged \acp{cme}. The red arrow marks the longitude of the later flare around 7\,UT.} \label{fig:scenarios} 
\end{figure}
As discussed above, the \ac{cme}-driven shock, which is still present during the appearance of the second component (see the type-II radio burst in Fig. \ref{fig:radio}), could explain the second component particle increases if its acceleration efficiency would suddenly increase for any reason.
However, it can hardly explain the very high-energy \acp{sep} only observed during the second component.\\
Another possible scenario for the second injection would be a particle trap which suddenly opens up and releases the second component \acp{sep}.
The sudden opening of this trap might be triggered by the solar activity around 7\,UT described above, explaining the temporal coincidence.
Thus we put forth the following scenario as a possible explanation (illustrated in Fig. \ref{fig:scenarios}) for the observed \ac{sep} event: A fraction of the particles accelerated early in the event is injected and later trapped in a closed magnetic region. 
The reason why only a fraction of the particles are trapped may be either due to a spatially extended injection (possibly formed by different accelerators such as the flare and the shock) where a part of the particles enters the open field lines, forming the first \ac{sep} component, and the second part enters the trap and is later released, forming the second component. 
The other possibility is a temporally extended injection where the first part of the particle injection reaches open field lines but the later accelerated part is injected into the trap because of the ongoing acceleration and the changing magnetic configuration.
The particle trap may be either a new magnetic structure, formed by the interacting  \ac{cme}1 and \ac{cme}2, or it may be the flux rope of one (or both) of the CMEs itself.
The shock driven by CME2 propagates through that region. 
Although shocks inside CMEs tend to have lower Mach numbers \citep{Lugaz2015} we expect further particle acceleration due to this scenario:
Similar to the twin-CME scenario proposed by \citet{Li2012} the shock driven by the second CME might experience an excess of seed population and an enhanced turbulence level increasing the efficiency of diffusive shock acceleration (first order Fermi acceleration). 
\citet{Kallenrode2001a, Kallenrode2001b} proposed this acceleration mechanism for events with large fluence where multiple shocks and \acp{cme} were involved  possibly forming particle mirrors. 
The region of the particle trap might also contain turbulence because of the interaction process of the two \acp{cme} and/or the passage of the shock through it \citep[e.g.,][]{Xiong2006}. Further stochastic acceleration (second order Fermi-acceleration) of the particles at this turbulence would therefore be plausible.
A steady leakage of the trapped particle population would explain the prolonged rising phase and anisotropy during the first part of the event.
However, ongoing acceleration and leakage driven by reconnection of the \ac{cme} flux tubes could also account for that \citep{Ruffenach2012}.
Indeed Fig. \ref{fig:var} shows that the interior of the two \acp{icme} observed at \ac{stb} was unusually disturbed as represented by the variances of the magnetic field RTN components.\\
We assume that the second component is released due to an opening of the particle trap. 
Around 8\,UT the interacting \acp{cme} have a height of $\sim$30-35 solar radii.
However, they are likely still magnetically anchored at the Sun.
Because of the temporal coincidence with the later activity around 7\,UT we suspect that it effects the \ac{cme}-\ac{cme} region in such a way that the particle-trap opening is triggered.
We note that when the particles are released from the trap they may again be further accelerated by the shock driven in front of the structure. 
The further rise of the \ac{sep} intensities for the following hours (see Fig. \ref{fig:intro_particles}), however, suggests that the particle trap is not completely destroyed, but that ongoing leakage happens.\\
This scenario is able to explain that the second component contains mainly high-energy particles: 
While no high energies were generated during the first phase of the event, particles were further accelerated inside the trap and released 4 hours later.
%
%
%
\section{Summary and Conclusion}\label{sec:summary}
The 26 Dec 2013 \ac{sep} event is a widespread electron and proton event observed at both STEREO and at L1 spacecraft (see Fig. \ref{fig:intro_particles}).
The STEREO spacecraft are separated by 59 degrees and span an angle of 210 degrees with L1 (from \ac{stb} over STA to the longitude of Earth, see Fig. \ref{fig:constellation}).
According to a Gaussian function applied to the peak intensities observed at the different positions (see Fig. \ref{fig:gaussians}) this event is, together with the 25 Feb 2014 event reported by \citet{Lario2016} and \citet{Klassen2016}, the widest event ever observed with the STEREO spacecraft: The Gaussian standard deviations of the peak intensity distributions for electrons yield 55 degrees (55-105\,keV) and 68 degrees (0.7-3\,MeV) and for 25-60\,MeV protons 58 degrees (see Fig. \ref{fig:gaussians}).
Usually the standard deviations vary between 30 and 50 degrees \citep{Lario2013, Dresing2014, Richardson2014}.
Because generally widespread electron events were observed relatively rarely compared to other \ac{sep} events \cite[less than 30 events during the STEREO era from 2007 to 2014, cf.][]{Dresing2014} it is likely that special conditions have to be present to produce these extraordinarily wide spreads. 
The main drivers under discussion are i) efficient perpendicular transport \citep{Droege2010, Droege2014, Droege2016, He2011, Dresing2012, Laitinen2013, Strauss2017}, ii) an extended acceleration region provided by a shock \citep{Lario2016}, and iii) an extended injection region \citep[cf.][]{Klein2008}.
Especially for electron events the role of an associated shock is not clear and whether or not it can efficiently accelerate solar energetic electrons is debated.\\
The 26 Dec 2013 event has a number of characteristics which require a spatial and temporal extended energetic electron (and proton) injection which might, at a first glance, be interpreted by ongoing acceleration by a shock.
These characteristics are:
\begin{itemize}
        \item Long-lasting 55-105\,keV electron anisotropies at all three observers, lasting up to about one day at \ac{stb} (see Fig. \ref{fig:intro_particles})
        \item Long-lasting rising phases of electron and proton intensities at all energies and observers (see Fig. \ref{fig:intro_particles})
        \item Small onset delays ($<30$ minutes) between the two STEREO spacecraft and the far-separated L1 spacecraft (see Fig. \ref{fig:intro_particles} and Table \ref{table:timeline})
        \item Very similar intensity time series observed at the two STEREO spacecraft, longitudinally separated by 59 degrees (see Fig. \ref{fig:intro_particles})
        \item The centers of the Gaussian distributions applied to the peak intensity observations are shifted westwards with respect to the associated flare location (see Fig. \ref{fig:gaussians})
\end{itemize}
The presence of at least two shocks (see type-II bursts in Fig. \ref{fig:radio2}) during the very beginning of the event and the very pronounced, later type-II radio burst extending far into the IP medium supports a shock scenario.
However, further distinct features of the event require a more complex situation than the simple shock scenario:
\begin{itemize}
        \item The centers of the Gaussian distributions applied to the intensity measurements at the same time (Fig. \ref{fig:running_gauss}) suggest that the source region of the near-relativistic electrons is an alternative source to that of the relativistic electrons and 25-60\,MeV protons: The centers of these higher-energy particles follow each other more closely and are situated more to the west, further away from the associated flaring \acp{ar}.\\
        We note that this point suggests that during the same \ac{sep} event, particles of different energy or species may not fill the IP space equally, which may be caused by different source regions or by transport effects.
        {\item A second component (later steepening of electron and proton intensity rises) is observed $\sim 4$ hours later than the first arriving particles, being more dominant at higher energies, which were partly not yet contained in the first component.}
\end{itemize}
As discussed in Sect. \ref{sec:second_comp} the presence of this second component points to additional acceleration and a new injection of \acp{sep} rather than simply to an intensification of the accelerator because of the different energy distributions of the two components.
The simple presence of a shock can therefore not explain these observations.\\
Hence we searched for further solar activity and found one potential candidate of a later solar flare occurring at the backside of the original event, which roughly fits temporally with the start of the second component.
However, we can exclude this flare as the parent source because of the relative spatial intensity distributions at the three observers (cf. Fig. \ref{fig:intro_particles}), which remain the same for both components suggesting that the source region of the second component is situated close to the one of the first component.
It is therefore suggested, that the source of the second \ac{sep} component must involve a trapping scenario.
Two \acp{cme} are associated to the complex solar activity which interact already before the first arriving particles are detected  (see Figs. \ref{fig:gcs} and \ref{fig:cme_interaction}). 
It is likely that the shock of the second \ac{cme} propagated through the preceding \ac{cme} and is then driven in front of the interacted structure. 
We put forth the following scenario for the 26 Dec 2013 event:
The \ac{cme}-interaction region might serve as a particle trap for flare and/or early shock-accelerated particles.
The region might contain turbulence either due to the interaction process or due to the passage of the shock of \ac{cme}2 through it. 
The unusually strong variances of the magnetic field components observed in-situ at \ac{stb} inside the \acp{icme} possibly reflect an enhanced turbulence and/or previous reconnection occurring inside the structure.
The further acceleration of the trapped particles may therefore have been driven by the shock passing through the structure, with enhanced acceleration due to the two \acp{cme} building a magnetic bottle configuration.
Alternatively, stochastic acceleration at the turbulence and/or at ongoing reconnection or simply the two converging \acp{cme} may have been involved.
The second component would then be caused by an opening of the trap which might be triggered by the later activity on the other side of the Sun.\\
We therefore conclude that the 26 Dec 2013 \ac{sep} event is caused by a complex combination of different mechanisms where the shock might play a role in the electron event. 
However, other important ingredients are the interaction of two \acp{cme} and a particle trap where further acceleration of energetic electrons and protons happens.

%
%
%
%
%
\begin{acknowledgements}
We acknowledge the STEREO PLASTIC, IMPACT, and SECCHI teams for providing the data used in this paper. 
The STEREO/SEPT Chandra/EPHIN and SOHO/EPHIN project is supported under grant 50OC1702 by the Federal Ministry of Economics and Technology on the basis of a decision by the German Bundestag. AV acknowledges the Austrian Science Fund FWF: P24092-N16. RGH acknowledges the financial support of the Spanish MINECO under project ESP2015-68266-R.
The NRH data survey is generated and maintained at the Observatoire de Paris by the LESIA UMR CNRS 8109 in cooperation with the Artemis team, Universities of Athens and Ioanina and the Naval Research Laboratory.
\end{acknowledgements}
\bibliographystyle{aa.bst}
\bibliography{references}
%
%
\end{document}